\documentclass[aps,prd,amssymb,amsmath,nobibnotes,nofootinbib,superscriptaddress, 12pt ]{revtex4}
\usepackage{amsfonts,amsmath,hyperref,url, color}
\allowdisplaybreaks
\usepackage{bm, bbm}
\usepackage{graphicx}
\usepackage{caption}
\usepackage{mathtools}
\usepackage{t1enc}
\usepackage{hepnames}
\usepackage{tensor}
\usepackage{comment}
\usepackage{float}
\usepackage{slashed}
\usepackage{cancel}
\usepackage{dcolumn}
\usepackage{mdframed}
\usepackage{braket}
\usepackage{dsfont}
\usepackage[normalem]{ulem}
\usepackage[dvipsnames]{xcolor}

\usepackage[normalem]{ulem}

\usepackage{adjustbox,booktabs}

\usepackage{makecell}
\usepackage{array}
\DeclareMathAlphabet\mathbfcal{OMS}{cmsy}{b}{n}

\newcommand{\bc}{\begin{center}}
\newcommand{\ec}{\end{center}}
\newcommand{\be}{\begin{eqnarray}}
\newcommand{\ee}{\end{eqnarray}}
\newcommand{\bs}{\begin{slide}}
\newcommand{\es}{\end{slide}}

\newcommand{\bi}{\begin{itemize}}
\newcommand{\ei}{\end{itemize}}

\newcommand{\ka}{\kappa}
\newcommand{\pd}{\partial}
\newcommand{\pdd}[1]{\frac{\pd}{\pd #1}}
\newcommand{\pdpd}[2]{\frac{\pd #1}{\pd #2}}
\newcommand{\xb}{\mathbf{x}}
\newcommand{\pb}{\mathbf{p}}
\newcommand{\qb}{\mathbf{q}}

\newcommand{\op}{\omega_{\mathbf{p}}}
\newcommand{\osp}{\omega_{S(\mathbf{p})}}

\newcommand{\oq}{\omega_{\mathbf{q}}}

\begin{document}
\title{$\kappa$-deformed spin-$1/2$ field}

\author{Tadeusz Adach}
\affiliation{University of Wroc\l{}aw, Faculty of Physics and Astronomy, pl.\ M.\ Borna 9, 50-204
Wroc\l{}aw, Poland}

\author{Andrea Bevilacqua}
\affiliation{National Centre for Nuclear Research, ul. Pasteura 7, 02-093 Warsaw, Poland}
\author{Jerzy Kowalski-Glikman}
\affiliation{National Centre for Nuclear Research, ul. Pasteura 7, 02-093 Warsaw, Poland}
\affiliation{University of Wroc\l{}aw, Faculty of Physics and Astronomy, pl.\ M.\ Borna 9, 50-204
Wroc\l{}aw, Poland}
\author{Giacomo Rosati}
\affiliation{Dipartimento di Matematica, Università di Cagliari, via Ospedale 72, 09124 Cagliari, Italy}
\affiliation{Istituto Nazionale di Fisica Nucleare, Sezione di Cagliari,
Cittadella Universitaria, 09042 Monserrato, Italy}
\author{Wojciech Wi\'slicki}
\affiliation{National Centre for Nuclear Research, ul. Pasteura 7, 02-093 Warsaw, Poland}

\date{\today}

\begin{abstract}
 In this paper, we investigate the Poincar\'e and discrete symmetries of a $\kappa$-deformed spin-$\tfrac12$ field, extending recent results obtained for scalar fields. We construct an action that is Poincar\'e invariant and analyze its consequences within the deformed framework. Our results confirm the findings of our recent analysis of the $\kappa$-deformed scalar field, where we established that there is no action invariant under both Poincar\'e symmetry and charge conjugation in the $\kappa$-deformed case, while $\mathcal{CPT}$-symmetry can be restored through a natural deformation of time reversal. Furthermore, we present an explicit calculation of the Noether charges associated with Poincar\'e symmetry and show that their algebra closes, demonstrating the internal consistency of the theory.

\end{abstract}

\maketitle

\section{Introduction}

The Effective Field Theory (EFT) paradigm (see, e.g., \cite{Burgess:2020tbq}) asserts that the predictions of a field theory can be reliably applied up to a specified energy scale, even if the full underlying theory governing higher-energy behavior remains unknown. This framework underpins our confidence in the predictions of the Standard Model, despite the expectation and necessity of physics beyond the Standard Model—for instance, to account for neutrino masses. According to EFT, physics below the Planck energy scale is governed by a local quantum field theory that assumes the Poincaré group as the spacetime symmetry group, provided the regions considered are sufficiently small to neglect general relativistic effects.

Quantum gravity, however, is expected to profoundly alter the structure of spacetime, rendering the effective field theory paradigm inapplicable. Nevertheless, it can be hypothesized that there exists a physical regime wherein quantum gravity effects lead to a replacement of local quantum field theory with a quantum-deformed field theory, with the deformation scale identified with the Planck energy scale (see \cite{Amelino-Camelia:2000stu}, \cite{Amelino-Camelia:2002cqb} for the original concept and \cite{Arzano:2021scz} for a recent review and detailed discussion). A potential realization of this hypothesis involves replacing the Poincaré algebra of Minkowski spacetime symmetries with its deformed Hopf algebra counterpart, the $\kappa$-Poincaré algebra \cite{Lukierski:1991pn}, \cite{Lukierski:1992dt}.

The $\kappa$-Poincaré algebra is the deformed symmetry of $\kappa$-Minkowski non-commutative spacetime \cite{Lukierski:1993wx}, \cite{Majid:1994cy}, and the momentum space of particles and/or field quanta exhibits curvature \cite{Kowalski-Glikman:2002oyi}, \cite{Kowalski-Glikman:2003qjp}. Models of particles and fields incorporating $\kappa$-deformation have emerged as some of the most widely used frameworks in quantum gravity phenomenology \cite{Amelino-Camelia:2008aez}, \cite{Addazi:2021xuf}, \cite{AlvesBatista:2023wqm}, \cite{Bevilacqua:2023swc}.

In a recent series of papers \cite{Arzano:2020jro}, \cite{Bevilacqua:2022fbz}, \cite{Bevilacqua:2024jpy}, \cite{Adach:2025pvj},\cite{CPTscalar} we conducted a detailed analysis of the theory of free complex scalar fields  in $\kappa$-Minkowski, based on a formulation of a $\kappa$-field theory that was originally developed in~\cite{Freidel:2007hk}. The $\kappa$-field theory model that was developed in~\cite{Freidel:2007hk} relies on a description of the $\kappa$-Poincar\'e symmetries\footnote{ Different formulations of field theories on $\kappa$-Minkowski are possible, which are based on different description of the $\kappa$-Poincar\'e symmetries, see for instance \cite{Lukierski:1992dt,Kosinski:1999ix,Amelino-Camelia:2001rtw,Agostini:2003vg,Meljanac:2010ps}.} in the ``classical basis''~\cite{LukKosMasClassicalBasis}. In that basis, the algebra of the symmetry generators is undeformed, while all the deformation lies in the coalgebra sector, which encodes the  action of the symmetry generators on multiparticle states, as discussed in \cite{Arzano:2022ewc} and references therein.

The principal outcome of the studies \cite{Arzano:2020jro}, \cite{Bevilacqua:2022fbz}, \cite{Bevilacqua:2024jpy}, \cite{Adach:2025pvj},\cite{CPTscalar} on the complex scalar field was the derivation of conserved Noether charges associated with the deformed $\kappa$-Poincaré algebra, and the characterization of the $\kappa$-deformed discrete symmetry sector.

In the papers \cite{Arzano:2020jro}, \cite{Bevilacqua:2022fbz}, \cite{Bevilacqua:2024jpy} our starting point was the action
\begin{align}
\mathcal{S}_{C}= & -\frac{1}{2}\int d^4x\left[\left(\partial_\mu \phi\right)^{\dagger}\star  \partial^\mu \phi+\left(\partial_\mu \phi\right) \star\left(\partial^\mu \phi\right)^{\dagger}
+m^2\, \left(\phi^{\dagger} \star \phi+\phi \star \phi^{\dagger}\right)\right] \label{scalar}
\end{align}
where the $\star$ encodes the noncommutative product arising from $\kappa$-Minkowski space. We  recall the properties of this $\star$-product in Appendix \ref{AppA}. We use $\dagger$ to denote a deformed conjugation, defined by its action on plane waves
\begin{align}\label{didag}
	\partial_i^\dag
	e^{-ipx}
	:=iS(\mathbf{p})_i
	e^{-ipx}
\,,\quad
	\partial_0^\dag
	e^{-ipx}
	:=iS(\omega_\mathbf{p})
	e^{-ipx}.
\end{align}
with $S$ being the antipode
\begin{align}
S\left(\omega_\mathbf{p}\right) & =-\omega_\mathbf{p}+\frac{\mathbf{p}^2}{\omega_\mathbf{p}+p_4}\,,\quad
S(\mathbf{p})  =-\frac{\kappa \mathbf{p}}{\omega_\mathbf{p}+p_4}, \quad  p_4 = \sqrt{\omega_\mathbf{p}^2 -\mathbf{p}^2+ \kappa^2}\label{antipode}
\end{align}
See \cite{Arzano:2020jro,Bevilacqua:2022fbz} for derivation and discussion. The action \eqref{scalar} is manifestly invariant under the (undeformed) $\mathcal{P}$ and $\mathcal{T}$ symmetries, and thanks to its form it is also $\mathcal C$-invariant. Explicitly, the charge conjugation action on the fields is defined as
\begin{align}
    \mathcal{C}^{-1} \phi \mathcal{C} = \phi^\dag\,,\quad \mathcal{C}^{-1} \phi^\dag \mathcal{C} = \phi
    \label{eq-cdef}
\end{align}
and the invariance of the action follows.

The action $\mathcal{S}_C$ can be viewed as the sum of two actions,
\begin{equation}
\mathcal{S}_1= - \int d^4x\left[\left(\partial_\mu \phi\right)^{\dagger}\star  \partial^\mu \phi
+ m^2\, \phi^{\dagger} \star \phi\right]
\label{S1scalar}
\end{equation}
\begin{equation}
\mathcal{S}_2= - \int d^4x\left[\left(\partial_\mu \phi\right) \star\left(\partial^\mu \phi\right)^{\dagger}
+ m^2\, \phi \star \phi^{\dagger}\right]
\label{S2scalar}
\end{equation}
which, taken individually, are Poincar\'e invariant, in the sense that they lead to conserved charges that, despite their forms differing from those in the undeformed theory, satisfy the  ordinary Poincar\'e algebra, in line with the classical basis of the $\kappa$-Poincaré generators.

 It turns out, as it was better understood in  \cite{Adach:2025pvj}, \cite{CPTscalar}, that the charges obtained from Noether analysis for the action $\mathcal{S}_C$ do not close an algebra, and Poincaré invariance is violated. This can be remedied by the introduction of a boundary term in $\mathcal{S}_C$ that restores Poincar\'e invariance, but at the cost of losing the invariance under charge conjugation. This is the strategy that was implicitly pursued in \cite{Arzano:2020jro}, \cite{Bevilacqua:2022fbz}, \cite{Bevilacqua:2024jpy} through the covariant phase space method.
The emerging picture (see \cite{CPTscalar}) is that in $\kappa$-field theory Poincar\'e invariance and ${\cal C}$ invariance are not compatible. As suggested in \cite{CPTscalar}, this can be understood as a consequence of the properties of the curved $\kappa$-momentum space.

One consequence of this is that under $\kappa$-deformation the charges associated with Poincar\'e transformations are not $\mathcal{C}$-invariant (although with the aforementioned boundary term it is possible to recover $\mathcal{C}$-symmetry for rotations).
 If, as argued in \cite{Arzano:2020jro}, parity and time-reversal discrete symmetries, $\mathcal{P}$ and $\mathcal{T}$, are not deformed, then $\mathcal{CPT}$ symmetry is broken at the level of charges.
Consequently, deviations from the standard $\mathcal{CPT}$ symmetry become increasingly pronounced with the momentum carried by the particles. The deformation disappears for particles at rest, and therefore the masses of particles and antiparticles are identical.

It was observed in~\cite{CPTscalar} that there is  an alternative natural definition of $\mathcal{CPT}$ symmetry for the deformed theory, in which $\mathcal{T}$ is modified in a way that complements the deformed charge conjugation, acting on the field as
\begin{equation}
\mathcal{T}^{-1}\phi\mathcal{T}=\phi^{\dagger*}(-t,\xb)
\end{equation}
Effectively, this swaps the momentum transformation properties of particles and antiparticles, restoring $\mathcal{CPT}$ as an exact symmetry of the theory. We refer to \cite{CPTscalar} for a thorough treatment of this approach. Yet another possible definition of discrete transformations was previously proposed  in \cite{Arzano:2016egk}, where the actions of $\mathcal{C}$, $\mathcal{P}$ and $\mathcal{T}$ were all modified by the deformation, but $\mathcal{CPT}$ invariance was lost. In light of this, we note that there is a certain degree of ambiguity in how discrete transformations are defined in the deformed theory, but charge conjugation symmetry appears to be fundamentally broken.

Having  understood the spin-0 free field, in this paper we  turn our attention to the spin-1/2 field. In the context of $\kappa$-deformation, spin-1/2 fields were first discussed in \cite{Lukierski:1993df} and then in modern language in \cite{Agostini:2002yd} (see also \cite{Harikumar:2009wv}, \cite{Andrade:2013oza}, \cite{Franchino-Vinas:2022fkh}).   Demanding Poincar\'e invariance turns out to restrict the form of the first order action to two possibilities, which correspond to generalizations of the two actions for scalar fields~(\ref{S1scalar}) or~(\ref{S2scalar}). We define the action for fermions as a generalization of~(\ref{S2scalar}), which amounts to a transposition of the usual fermion Lagrangian (see Sec.~\ref{sec:deformed spin theory}).

It is easy to see that the Dirac equation we consider
coincides with the one that was derived in~\cite{Agostini:2002yd}.
This is because the main result of~\cite{Agostini:2002yd} was to prove that to have a $\kappa$-deformed Dirac equation based on standard, energy-independent, $\gamma$ matrices, the equation must be defined in terms of the derivatives belonging to the 5D differentiation calculus \cite{Sitarz:1994rh}, \cite{deAzcarraga:1995uw}, \cite{Freidel:2007hk}.
The 5D calculus is precisely the starting point of the construction we have made in the previous works \cite{Arzano:2020jro}, \cite{Bevilacqua:2022fbz}, \cite{Bevilacqua:2024jpy},\cite{CPTscalar},
and on which this work is based. The present paper builds on and expands the result of~\cite{Agostini:2002yd} by connecting the deformed Dirac equation to an explicit action, as well as examining its Poincar\'e and discrete symmetries.

The plan of the paper is as follows. In the following Sec.~\ref{sec:undeformed spin theory} we introduce the Lagrangian of the spin-1/2 field, discuss its properties and briefly recall how its symplectic structure, the associated Noether charges and their algebra are obtained, which we will then directly generalize to the deformed case. In Sec.~\ref{sec:deformed spin theory} we discuss the form of the deformed spin-1/2 Lagrangian and then compute the deformed Noether charges. Here the calculations are considerably more elaborate than in the undeformed case and we move most of the details to Appendix \ref{AppB}. We also derive the deformed symplectic structure and show that the charges obtained in Sec.~\ref{sec:deformed spin theory} indeed satisfy the Poincar\'e algebra, as they should.  Sec.~\ref{sec:cpt-both} is devoted to discrete symmetries (or lack thereof) of the action and charges examined in the previous sections. We examine two prescriptions: one where $\mathcal C$, $\mathcal P$ and $\mathcal T$ are defined exactly as in the undeformed theory and one based on the approach developed in~\cite{CPTscalar} that is tailored to ensure $\mathcal{CPT}$ invariance.  Finally, in Sec.~\ref{sec:conclusions} we summarize the results obtained and briefly discuss the  possible phenomenological consequences of the theory of deformed spin-1/2 fields.

\section{Undeformed spin-1/2 action, Noether charges, and symplectic form}
\label{sec:undeformed spin theory}

In this section we recall the spin-1/2 Dirac theory using the techniques which will prove convenient to discuss deformed generalization of this theory. Our starting point is the Dirac Lagrangian
\begin{align}\label{LDirac}
	\mathcal{L}
	=
	\bar{\psi}\left(i
	\slashed{\partial}
	-
	m\right)\psi
\end{align}
with $\bar\psi = \psi ^\dag\gamma^0$. As we will explicitly show in a moment, this Lagrangian is Poincar\'e-invariant. It is also invariant under the discrete spacetime symmetries $\mathcal{C}$, $\mathcal{P}$, $\mathcal{T}$, which we will discuss in detail below, in Sect.\ 4.

To begin the analysis, we calculate the variation of the action:
\begin{align}
    \delta \mathcal{S}&=
    \int d^{4}x\,    \left[  \delta\bar{\psi}    \left(i\slashed\pd-m\right)\psi+\bar{\psi}\left(i\slashed\pd-m\right)\delta\psi\right]=\nonumber\\
    &=\int d^{4}x\,\left[\delta\bar{\psi}\left(i\slashed\pd-m\right)\psi-\bar{\psi}\left(i\overleftarrow{\slashed{\pd}}+m\right)\delta\psi+\pd_{\mu}\left(i\bar{\psi}\gamma^{\mu}\delta\psi\right)\right],\label{eq:vars}
\end{align}
from which we immediately read off the equations of motion:
\begin{align}
\left(i\slashed\pd-m\right)\psi=0\,,\quad
\bar{\psi}\left(i\overleftarrow{\slashed\pd}+m\right)=0\label{eq:eom-psib}
\end{align}
Inserting the mass-shell condition $p^2=m^2$ into the Fourier transform, we obtain the on-shell field
\begin{align}
\psi(x)&=\int d^4 p\,\delta(p^2-m^2)\tilde\psi(p)e^{-ipx}=\int \frac{d^3 p}{\sqrt{2\op}} \left[\mathfrak{u}(\pb)e^{-ipx}+\mathfrak{v}(\pb )e^{ipx}\right]\
\label{eq:psi-modexp}
\end{align}
where $\op=\sqrt{\pb^2+m^2}$. It immediately follows that
\begin{align}
\bar\psi(x)=\int \frac{d^3 p}{\sqrt{2\op}} \left[\bar{\mathfrak{u}}(\pb)e^{ipx}+\bar{\mathfrak{v}}(\pb )e^{-ipx}\right]\label{eq:psib-modexp}
\end{align}
The resulting momentum space field equations
\begin{align}\label{eq:eom-p}
    \slashed{p} \mathfrak{u} - m  \mathfrak{u} = 0 \,,\quad  \slashed{p} \mathfrak{v} + m  \mathfrak{v} =0
\end{align}
can be solved explicitly if one chooses a gamma-matrices representation. In Dirac representation
\begin{align}
    \gamma^0
	=
	\begin{pmatrix}
		1 & 0 \\
		0 & -1
	\end{pmatrix} \,,\quad
    \gamma^j
	=
	\begin{pmatrix}
		0 & \sigma^j \\
		-\sigma^j & 0
	\end{pmatrix} \,,\quad
\end{align}
it is easy to check that  general solution is given by\footnote{The position of the spin label $s$ is  not meaningful, but we adopt this notation for visual clarity and to stress summation.}
\begin{align}
    \mathfrak{u}(\mathbf{p}) &= a_\mathbf{p}^s \, u_s(\pb) \,,\quad \mathfrak{v}(\mathbf{p}) = b^{\dag s}_\mathbf{p} \, v_s(\pb)\nonumber\\
    \bar{\mathfrak{u}}(\mathbf{p}) &= a^{\dag s}_\mathbf{p} \, \bar{u}_s(\pb) \,,\quad \bar{\mathfrak{v}}(\mathbf{p}) = b_\mathbf{p}^s \, \bar{v}_s(\pb)
\end{align}
\begin{align}
	u_s(\mathbf{p})
	=\sqrt{\op+m}
	\begin{pmatrix}
		 \xi_s \\
		\frac{\pb\cdot\sigma}{\op+m} \xi_s
	\end{pmatrix}
	\qquad
	v_s(\mathbf{p})
	=\sqrt{\op+m}
	\begin{pmatrix}
		\frac{\pb\cdot\sigma}{\op+m} \eta_s \\
		 \eta_s
	\end{pmatrix}
    \label{eq:uv-def}
\end{align}
with $\xi_s$, $\eta_s$ being two (possibly different) 2-dimensional basis vectors satisfying
\begin{align}\label{xinorm}
	\xi^{\dag}_s \xi_{s'} = \delta_{ss'}
	\qquad
	\eta^{\dag}_s \eta_{s'} = 	\delta_{ss'}
	.
\end{align}
After these preliminaries let us turn to the discussion of symmetries of undeformed spin-1/2 theory.

In order to compute the conserved charges of the theory using the canonical method, we impose the field equations  \eqref{eq:eom-psib},
which results in the equation
\begin{align}
\delta\mathcal{L}=\pd_\mu\left(i\bar\psi\gamma^\mu \delta\psi\right)\label{eq:vars-noether}
\end{align}
Note that since the Lagrangian of spin-1/2 field vanishes on-shell, $\mathcal{L}=\bar\psi\times EOM$, the left-hand side automatically vanishes when equations of motion (EOM) are satisfied. Plugging in the translation transformation
\begin{align}
\delta_\mathrm{T}\psi=\epsilon^\nu \pd_\nu \psi \,,\quad
\delta_\mathrm{T}\bar\psi=\epsilon^\nu \pd_\nu \bar\psi,\label{eq:vart-psi-u}
\end{align} we obtain the continuity equation
\begin{align}
\epsilon^{\nu}\pd_{\mu}\left(i\bar{\psi}\gamma^{\mu}\pd_{\nu}\psi\right)=0.
\end{align}
We identify the Noether current for translations with the energy-momentum tensor
\begin{align}
\pd_{\mu}\left(i\bar{\psi}\gamma^{\mu}\pd_{\nu}\psi\right)=\pd_\mu T^\mu_{\,\,\nu}=0.
\end{align}
The translation charge is then the integral over space of the time component of the current
\begin{align}
\mathcal{P_\mu}=\int d^3x\,\,T^0_{\,\,\mu}=\int d^3x\,\,i\psi^\dag \pd_{\mu}\psi
\end{align}
Plugging in the field expansions \eqref{eq:psi-modexp} and \eqref{eq:psib-modexp}, and using the identities
\begin{align}
u_{s}^{\dagger}(\pb)u_{s'}(\pb)=v_{s}^{\dagger}(\pb)v_{s'}(\pb)&=2\op\delta_{ss'}\,,\quad
u_{s}^{\dagger}(\pb)v_{s'}(-\pb)=v_{s}^{\dagger}(-\pb)u_{s'}(\pb)=0\label{eq:uv}
\end{align}
we derive
\begin{align}\label{transation-u}
\mathcal{P}_\mu &= \int d^{3}p\,\,p_{\mu}\left(a_{\pb}^{\dagger s}a_{\pb}^{s}+b_{\pb}^{\dagger s}b_{\pb}^{s}\right)
\end{align}

For the Lorentz sector, the infinitesimal variations contain two contributions, the `orbital', identical to the case of a scalar field and the `spinorial', rotating the spinor's components,
\begin{align}
\delta_\mathrm{L}\psi=\omega_{\mu\nu}\left(x^\mu \pd^\nu\psi-iS^{\mu\nu}\psi\right)\,,\quad
\delta_\mathrm{L}\bar\psi=\omega_{\mu\nu}\left(x^\mu\pd^\nu\bar\psi+i\bar\psi S^{\mu\nu}\right)
\end{align}
where
\begin{align}
S^{\mu\nu}=\frac i 4 \left[\gamma^\mu,\gamma^\nu\right]. \qquad \omega_{\mu\nu}=-\omega_{\nu\mu}.
\end{align}

For rotations, the variation of the field is
\begin{align}
\delta_{M_k}\psi\equiv\epsilon^{ijk}\left(x^i\pd^j+\frac{1}{4}\gamma^i\gamma^j\right)\psi
\end{align}
and the Noether charge associated with rotations evaluates to
\begin{align}\label{rotation-u}
\mathcal{M}^k=\int d^{3}p\,\,\left[i\epsilon^{ijk}p^{i}\left(a_{\pb}^{\dagger s}\pdpd{a_{\pb}^{s}}{p_{j}}+b_{\pb}^{\dagger s}\pdpd{b_{\pb}^{s}}{p_{j}}\right)+\frac{1}{2}\sigma^k_{ss'}\left(a_{\pb}^{\dagger s}a_{\pb}^{s'}+b^{\dagger s}_{\pb }b_{\pb }^{s'} \right)\right]
\end{align}

For boosts, we have the variation
\begin{align}
\delta_{N_j}\psi\equiv \left[-\left(x^0\pd^j-x^j\pd^0\right)+\frac{1}{2}\gamma^0\gamma^j\right]\psi
\label{eq:varpsi-n}
\end{align}
and the boost charge is expressed by
\begin{align}\label{boost-u}
\mathcal{N}^{j}=\int d^{3}p\,\left[-\frac{i}{2}\op\left(a_{\pb}^{\dagger s}\pdpd{a_{\pb}^{s}}{p_{j}}-\pdpd{a_{\pb}^{\dagger s}}{p_{j}}a_{\pb}^{s}+b_{\pb}^{\dagger s}\pdpd{b_{\pb}^{s}}{p_{j}}-\pdpd{b_{\pb}^{\dagger s}}{p_{j}}b_{\pb}^{s}\right)+\frac{1}{2}\frac{\epsilon^{jkl}p^{k}\sigma_{ss'}^{l}}{\op+m}\left(a_{\pb}^{\dagger s}a_{\pb}^{s'}+b_{\pb}^{\dagger s}b_{\pb}^{s'}\right)\right]
\end{align}

In order to obtain the symplectic form, we only need to take the second variation of the time component of the presymplectic current, which we have already calculated in \eqref{eq:vars-noether}
\begin{align}
\Omega=\;\int_{\Sigma_{t}}d^{3}x\delta\left(i\bar{\psi}\gamma^{0}\delta\psi\right)=i\int_{\Sigma_{t}}d^{3}x\left(\delta\psi^{\dagger}\wedge\delta\psi\right)=i\int_{\Sigma_{t}}d^{3}p\left(\delta a_{\pb}^{\dagger s}\wedge\delta a_{\pb}^{s}+\delta b_{\pb}^{s}\wedge\delta b_{\pb}^{\dagger s}\right)
\end{align}
The spin-statistics theorem (see e.g., \cite{Srednicki:2007qs}) tells that spin 1/2-fields must be anti-commuting, so that $\delta b_{\pb}^{s}\wedge\delta b_{\pb}^{\dagger s} =\delta b_{\pb}^{\dagger s}\wedge\delta b_{\pb}^{ s}  $ and the fundamental anti-commutators take the form
\begin{align}\label{acom-u}
\left[ a_{\pb}^{s},a_{\qb}^{\dagger s'}\right]_+ =\delta^{3}(\pb-\qb)\delta^{ss'}\qquad \left[ b_{\pb}^{s},b_{\qb}^{\dagger s'}\right]_+ =\delta^{3}(\pb-\qb)\delta^{ss'}
\end{align}
with all other anti-commutators vanishing. From now on we will call $a_{\pb}^{s}$, $a_{\pb}^{\dagger s'}$, $b_{\pb}^{s}$, and $b_{\pb}^{\dagger s'}$ the particle and antiparticle creation and annihilation operators, respectively. Applying the fundamental anti-commutators \eqref{acom-u}, one can show through a somewhat tedious, but textbook calculation that the charges \eqref{transation-u}, \eqref{rotation-u} and \eqref{boost-u} indeed satisfy the Poincar\'e algebra.

\section{Deformed spin-1/2 theory}
\label{sec:deformed spin theory}

The $\kappa$-deformed theory differs from the undeformed one in two major respects. First, the standard commutative product of functions of the undeformed theory is replaced with the non-commutative star product, reflecting the non-commutativity of deformed spacetime. Second, the commutative composition of momenta $p+q$ is replaced by the non-commutative composition rule denoted by $p\oplus q$ and as a consequence also the inverse momentum is denoted by $\ominus p$ or $S(p)$ (in what follows, we will mainly use the notation $S(p)$). The reader can consult Appendix \ref{AppA} and \cite{Arzano:2021scz} for more details.

The first question to ask is what is the form of the deformed Lagrangian. Up to integration by parts, there are (naively) four possible distinct\footnote{One should note that under the noncommutative product the bilinears $\bar\psi\star\psi$ and $-\psi^T\star\bar\psi^T$ are not equivalent.} deformed analogs of \eqref{LDirac}:
\begin{equation}\label{k-lagrangians}
\begin{aligned}
\mathcal L^\kappa_1&=\bar\psi\star\left(i\gamma^\mu\pd_\mu-m\right)\psi,&\qquad \mathcal L^\kappa_2&=\bar\psi\star\left(-i\gamma^\mu S(\pd_\mu)-m\right)\psi,\\
\mathcal{L}_3^{\kappa}&=-\psi^{T}\left(i\gamma^{\mu T}\overleftarrow{\pd_{\mu}}-m\right)\star\bar{\psi}^{T},&\qquad \mathcal{L}_4^{\kappa}&=-\psi^{T}\left(-i\gamma^{\mu T}\overleftarrow{S(\pd_{\mu})}-m\right)\star\bar{\psi}^{T}
\end{aligned}
\end{equation}
However, the asymmetric nature of derivative coproducts inherited from the translation sector of $\kappa$-Poincar\'e \eqref{DeltaP0}-\eqref{DeltaPi} provides a natural restriction to only two of the choices listed above\footnote{This structural obstacle was not previously encountered when working with the scalar field, since the (deformed) Klein-Gordon operator is quadratic in derivatives.}.
To demonstrate this, we examine the coproduct of spatial derivatives of two arbitrary functions $\phi(x)$ and $\varphi(x)$ that we will make use of when integrating by parts:
\begin{equation}
\pd_i(\phi\star\varphi)=\pd_i\phi\star\frac{\Delta_+}{\ka}\varphi+\phi\star\pd_i\varphi
\end{equation}
which leads to the following deformed Leibniz rules:
\begin{equation}\label{pd-left}
\pd_i\phi\star\varphi=\pd_i\left(\phi\star\xi\right)+\phi\star S(\pd_i)\xi
\end{equation}
where $\xi=\ka\Delta_+^{-1}\varphi$, and
\begin{equation}\label{pd-right}
\begin{aligned}
\phi\star\pd_i\varphi&=\pd_{i}(\phi\star\varphi)-\pd_{i}\phi\star\frac{\Delta_{+}}{\ka}\varphi\\
&=\pd_{i}(\phi\star\varphi)+\frac{i}{\ka}\pd_{0}\left(S(\pd_{i})\phi\star\varphi\right)+\frac{i}{\ka}\pd_{4}\left(S(\pd_{i})\phi\star\varphi\right)+\left(S(\pd_{i})\phi\star\varphi\right)
\end{aligned}
\end{equation}
We can see that, due to the presence of the total time derivative term in \eqref{pd-right}, for kinetic terms in the Lagrangian of the form~\eqref{pd-right}, that is $\bar\psi\star\slashed{\pd}\psi$ or $S(\slashed\pd)\psi\star\bar\psi$, the integration by parts with respect to the spatial derivatives will produce an additional boundary term on the constant time surface, which will contain the $\gamma^i$ matrix (notice that the standard boundary terms at the constant time surface contain only expressions of the form $\bar\psi \gamma^0\psi$.)  It can be shown that if such a term is present, the canonical Noether charges are not time-independent, which signals the manifest breaking of spacetime symmetries. As a consequence, the $\kappa$-deformed actions
 $\mathcal L^\ka_1$ and $\mathcal L^\ka_4$ are not $\kappa$-Poincar\'e invariant and we disregard them in what follows.  We are therefore left with $\mathcal L^\ka_2$ and $\mathcal L^\ka_3$, which both describe  well-behaved theories, and their mutual relation will be elaborated on in later sections. For the remainder of this section, we choose to work with $\mathcal L^\ka_3$, as it is most reminiscent of the undeformed Lagrangian, and we denote
\begin{align}\label{defLDirac}
\mathcal{L}^{\kappa}=-\psi^{T}\left(i\gamma^{\mu T}\overleftarrow{\pd_{\mu}}-m\right)\star\bar{\psi}^{T}\,,\quad S^\kappa = \int d^4x \mathcal{L}^{\kappa}.
\end{align}

From the variation of the action \eqref{defLDirac}
\begin{align}
\delta \mathcal{S}^\kappa =&\;-\int d^4 x\left[\delta\psi^{T}\star\left(i\gamma^{\mu T}S(\pd_{\mu})-m\right)\bar{\psi}^{T}+\psi^{T}\left(i\gamma^{\mu T}\overleftarrow{\pd_{\mu}}-m\right)\star\delta\bar{\psi}^{T}+\text{surface terms}\right]\label{eq:vars-k3}
\end{align}
we obtain the equations of motion
\begin{align}
\left(i\gamma^{\mu T}S(\pd_{\mu})-m\right)\bar{\psi}^{T}=0\,,\quad
\psi^{T}\left(i\gamma^{\mu T}\overleftarrow{\pd_{\mu}}-m\right)=0\label{eq:eom-psi-k3}
\end{align}
which are equivalent to
\begin{align}
\left(i\slashed{\pd}-m\right)\psi&=0\label{defEOMpart}\\
\bar\psi\left(i\overleftarrow{S(\slashed{\partial})}-m\right)&=0\label{defEOMapart}
\end{align}
The on-shell relations following from these equations are  undeformed thanks to the fact that $p_\mu p^\mu=S(p_\mu)S(p^\mu)$. The solutions of the equations \eqref{defEOMpart} and \eqref{defEOMapart} differ, however, from the solutions of the undeformed equations. To see this, let us proceed with the mode expansion. As in the undeformed case, the on-shell spinors can be expressed in terms of  $\mathfrak{u}(\pb )$ and $\mathfrak{v}(\pb)$, as follows
\begin{align}
\psi^T(x)
&=\int\frac{d^{3}p}{\sqrt{2\op}p_{4}/\ka}\left[u^T_s(\pb ) a^s_\pb e^{-ipx}+v^T_s(-S(\pb) )b^{\dagger s}_{\pb}e^{-iS(p)x}\right]\label{eq:modexp-psi-k3}\\
\bar{\psi}^T(x)
&=\int\frac{d^{3}p}{\sqrt{2\op}p_{4}/\ka}\left[\bar{u}^T_s(\pb ) a^{\dagger s}_\pb e^{-iS(p)x}+\bar{v}^T_s(-S(\pb) )b^{ s}_{\pb}e^{-ipx}\right]\label{eq:modexp-psib-k3}
\end{align}
where we wrote the expressions for transposed spinors, which we will find convenient when we turn to the calculation of Noether charges. In the formulas above
\begin{align}
u_{s}^{T}(\mathbf{p})=\sqrt{\op+m}\left(\begin{array}{cc}
\xi_{s}^{T} & \xi_{s}^{T}\frac{\pb\cdot\sigma}{\op+m}\end{array}\right)
	\qquad
	v_{s}^{T}(-S(\mathbf{p}))=\sqrt{\osp+m}\left(\begin{array}{cc}
\eta_{s}^{T}\frac{-S(\pb)\cdot\sigma}{\osp+m} & \eta_{s}^{T}\end{array}\right)
    \label{eq:uv-defk3}
\end{align}
and the spinorial identities look now as follows
\begin{align}
u_{s}^{T}(\pb)u_{s'}^{*}(\pb)&=u_{s'}^{\dagger}(\pb)u_{s'}(\pb)=2\op\delta_{ss'} \\ v_{s}^{T}(-S(\pb))v_{s'}^{*}(-S(\pb))&=v_{s'}^{\dagger}(-S(\pb))v_{s}(-S(\pb))=2\osp\delta_{ss'}\\
v_{s}^{T}(-S(\pb))u_{s'}^{*}(S(\pb))&=u_{s'}^{\dagger}(S(\pb))v_{s}(-S(\pb))=0\label{eq:vuk3}\\ u_{s}^{T}(S(\pb))v_{s'}^{*}(-S(\pb))&=v_{s'}^{\dagger}(-S(\pb))u_{s}(S(\pb))=0
\label{eq:uvk3}
\end{align}
After these preliminaries we can turn to the analysis of the spacetime symmetries of the action and the associated Noether charges.
\subsection{Variation of the action}
To proceed, we consider the variation $\delta\mathcal{L}^\kappa$ and assume that the equations of motion are satisfied. Using the fact that
$\ka\Delta_+^{-1}-1=i\ka^{-1}S(\pd_0)+i\ka^{-1}S(\pd_4)$ and that $S(\pd_4)=\pd_4$ and making use of \eqref{eq:vars-k3} we find
\begin{align}
\delta\mathcal{L}^\kappa=&\;-\pd_{0}\left(\delta\psi^{T}\star i\gamma^{0T}\ka\Delta_{+}^{-1}\bar{\psi}^{T}\right)-\pd_{j}\left[\left(\delta\psi^{T}\star i\gamma^{jT}\ka\Delta_{+}^{-1}\bar{\psi}^{T}\right)+i\ka\Delta_{+}^{-1}\left(\delta\psi^{T}\star i\gamma^{0T}\Delta_{+}^{-1}\pd^{j}\bar{\psi}^{T}\right)\right]\nonumber\\
&-\left(i\ka^{-1}S(\pd_{0})+i\ka^{-1}\pd_{4}\right)\left(\delta\psi^{T}\star i\gamma^{0T}S(\pd_{0})\bar{\psi}^{T}\right)
\end{align}

We now substitute the explicit form of $S(\pd_0)=-\pd_0-i\Delta_+^{-1}\pd_i\pd^i$:
\begin{align}
\delta\mathcal{L}^\kappa=&\;-\pd_{0}\left(\delta\psi^{T}\star i\gamma^{0T}\ka\Delta_{+}^{-1}\bar{\psi}^{T}\right)-\pd_{j}\left[\left(\delta\psi^{T}\star i\gamma^{jT}\ka\Delta_{+}^{-1}\bar{\psi}^{T}\right)+i\ka\Delta_{+}^{-1}\left(\delta\psi^{T}\star i\gamma^{0T}\Delta_{+}^{-1}\pd^{j}\bar{\psi}^{T}\right)\right]\nonumber\\
&-\pd_{0}\left(\delta\psi^{T}\star\ka^{-1}\gamma^{0T}S(\pd_{0})\bar{\psi}^{T}\right)-\pd_{j}\left[i\Delta_{+}^{-1}\pd^{j}\left(\delta\psi^{T}\star\ka^{-1}\gamma^{0T}S(\pd_{0})\bar{\psi}^{T}\right)\right]\nonumber\\
&+\pd_{4}\left(\delta\psi^{T}\star\ka^{-1}\gamma^{0T}S(\pd_{0})\bar{\psi}^{T}\right)\nonumber\\
&=\;-\pd_{0}\left(\delta\psi^{T}\star\Pi^{0}\right)-\pd_{j}\Phi^{j}-\pd_{4}\left(\delta\psi^{T}\star\Pi^{4}\right)\label{eq:vars-noether-k3}
\end{align}
where, considering that $\gamma^{0T}\bar\psi ^T=\psi^*$,
\begin{align}
\Pi^{0}&=\left(i\ka\Delta_{+}^{-1}+\ka^{-1}S(\pd_{0})\right)\psi^{*}\\
\Pi^{4}&=-\ka^{-1}S(\pd_{0})\psi^{*}\\
\Phi^{j}&=\left(\delta\psi^{T}\star i\gamma^{jT}\ka\Delta_{+}^{-1}\bar{\psi}^{T}\right)+i\ka\Delta_{+}^{-1}\left(\delta\psi^{T}\star i\Delta_{+}^{-1}\pd^{j}\psi^{*}\right)+i\Delta_{+}^{-1}\pd^{j}\left(\delta\psi^{T}\star\ka^{-1}S(\pd_{0})\psi^{*}\right)\label{eq:phij}
\end{align}
Let us take a closer look at the conjugate momentum $\Pi^0$:
\begin{align}
\Pi^0&=\left(i\ka\Delta_{+}^{-1}-\ka^{-1}S(\pd_{0})\right)\psi^{*}=\left(-\ka^{-1}\pd_{4}+i\right)\psi^{*}=i\frac{i\pd_{4}+\ka}{\ka}\psi^{*}
\end{align}
Upon Fourier transform, $i\partial_4+\kappa$ becomes $ p_4$, which on-shell is $p_4=\sqrt{\ka^2+m^2}$ (a number), thus $\Pi^0$ is proportional to the undeformed momentum $i\psi^\dagger$.
\subsection{Translations}

In the active picture, the field transforms under infinitesimal translations as
\begin{align}
\delta_\mathrm{T}\psi=&\epsilon^A \pd_A \psi\,,\quad \delta_\mathrm{T}\psi^T=\epsilon^A \pd_A \psi^T \label{eq:vart-psi}\\
\delta_\mathrm{T}\bar\psi=&\epsilon^A \pd_A \bar\psi \,,\quad \delta_\mathrm{T}\bar\psi^T=\epsilon^A \pd_A \bar\psi^T \label{eq:vart-psib}
\end{align}
In the formulas above we took the liberty of extending spacetime translations by the `translation' in the direction $4$. This comes naturally from the fact that we are here adopting\footnote{See~\cite{OecklDiffCalc,Rosati:2021sdf} for a discussion on the properties of the alternative 4-dimensional calculus on $\kappa$-Minkowski, that was at the basis of other works on $\kappa$-field theories~\cite{Agostini:2006nc}.} the 5-dimensional (bi)-covariant differential calculus on $\kappa$-Minkowski space defined in \cite{Sitarz:1994rh}, \cite{deAzcarraga:1995uw}, \cite{Freidel:2007hk}.
Moreover, we should take into account that the change of the fields due to translations obeys the Leibniz rule with respect to the star product:
\begin{align}
\delta_\mathrm{T}(\phi\star\psi)=\delta_\mathrm{T}\phi\star\psi+\phi\star\delta_\mathrm{T}\psi\label{eq:dt-leibniz}
\end{align}
This is because the analogous transformation in non-commutative space is the derivation $d=d\hat x^A\hat\partial_A$, which satisfies Leibniz rule by definition. In turn, the parameters of transformations $\epsilon^A$ do not commute with noncommutative $\kappa$-Minkowski coordinates \cite{Sitarz:1994rh,Freidel:2007hk,Arzano:2020jro}.

To calculate the translation charge, we put in the translational variation $\delta_\mathrm{T}$ into \eqref{eq:vars-noether-k3}:
\begin{align}
\delta_\mathrm{T}\mathcal{L}^\kappa=-\pd_{0}\left(\delta_\mathrm{T}\psi^{T}\star\Pi^{0}\right)-\pd_{j}\Phi^{j}_T-\pd_{4}\left(\delta_\mathrm{T}\psi^{T}\star\Pi^{4}\right)\label{eq:current-k3}
\end{align}

Now, as in the undeformed case, we notice that the Lagrangian $\mathcal{L}^\kappa$ has the form $EOM\star{\bar\psi}^T$ and therefore it vanishes on shell. As a consequence, we get the continuity equation
\begin{align}
-\pd_{0}\left(\pd_B\psi^{T}\star\Pi^{0}\right)-\pd_{j}\Phi^{j}_B-\pd_{4}\left(\pd_B\psi^{T}\star\Pi^{4}\right)=0\label{eq:noether-k3}
\end{align}
The translation charge is then, as in the undeformed case, the integrated time component of the current (energy-momentum tensor):
\begin{align}\label{momdef}
\mathcal{P_\mu^\kappa}=-\int d^3x\;\pd_\mu\psi^{T}\star\Pi^{0}=\int\frac{d^{3}p}{p_{4}/\ka}\left[\frac{p_{+}^{3}}{\ka^{3}}p_{\mu}a_{\pb}^{\dagger s}a_{\pb}^{s}-S(p_{\mu})b_{\pb}^{\dagger s}b_{\pb}^{s}\right]
\end{align}
We see that~\eqref{momdef} is time-independent, which proves that it is a conserved charge associated with a symmetry of the theory.
There is also a conserved charge associated with the derivative $\partial_4$ which reads
\begin{align}
\mathcal{P}^\kappa_4=\int\frac{d^{3}p}{p_{4}/\ka}\left[\frac{p_{+}^{3}}{\ka^{3}}a_{\pb}^{\dagger s}a_{\pb}^{s}-b_{\pb}^{\dagger s}b_{\pb}^{s}\right](p_4-\ka)
\end{align}
It can be shown that the Noether charge $\mathcal{P}^\kappa_4$ is proportional to the electric charge carried by the particle~\cite{Arzano:2009ci}.

The details of the computations are presented in Appendix \ref{AppB}.

\subsection{Lorentz transformations}
As it was shown in \cite{Freidel:2007hk}, the Lorentz action on noncommutative plane waves can be expressed as
\begin{align}
N_j\triangleright\hat e_k&=i\left(\hat x_jP_0(k)-\hat x_0P_j(k)\right)e^{-\frac{k_0}{\ka}}\hat e_k\\
M_l\triangleright\hat e_k&=i(\epsilon_{ijl}P_i(k)\hat x_j)e^{-\frac{k_0}{\ka}}\hat e_k
\end{align}
Applying the Weyl map (see Appendix \ref{AppA}) and noticing that $e^{-k_0/\ka}=\ka P_+^{-1}$, we obtain the Lorentz action in $\star$-product formalism
\begin{align}
N_j\triangleright e^{ipx}&=x_{[0}\star\frac{\kappa}{\Delta_+}\pd_{j]}e^{ipx}\label{boost-epx-def}\\
M_l\triangleright e^{ipx}&=\epsilon_{ijl}x_{[i}\star\frac{\kappa}{\Delta_+}\pd_{j]}e^{ipx}\label{rotation-epx-def}
\end{align}
Since the $\gamma$ matrix structure is unaffected by the deformation~\cite{Agostini:2002yd}, the spinorial contribution is the same as in the undeformed case, while the orbital part is constructed from~\eqref{boost-epx-def}-~\eqref{rotation-epx-def}. The infinitesimal change of the field under a Lorentz transformation thus takes the form
\begin{align}
\delta_\mathrm{L}\psi=&\omega_{\mu\nu}\left(x^\mu \star\frac{\ka}{\Delta_+}\pd^\nu\psi-iS^{\mu\nu}\psi\right)
\end{align}
where the parameter $\omega_{\mu\nu}=-\omega_{\nu\mu}$ is again such that $\delta_L$ is a derivation with respect to the $\star$-product. As we will see, the charges associated with this transformation are conserved (time independent) which proves that the transofrmations they generate are symmetries of the action. Along with the translational charges derived above, they form the Poincar\'e algebra. Since for $\mathcal{L}^\kappa$ the on-shell variation is the boundary term, we can again use a conservation equation analogous to \eqref{eq:current-k3}, which we used to obtain translational charge
\begin{align}
-\pd_{0}\left(\delta\psi^{T}\star\Pi^{0}\right)-\pd_{j}\Phi^{j}-\pd_{4}\left(\delta\psi^{T}\star\Pi^{4}\right)=0\label{eq:noether-k3-2}
\end{align}

\subsubsection{Rotation charge}
Following the undeformed case, we postulate
\begin{align}
\delta_{M_k}\psi^T=\epsilon^{ijk}\left(x^{i}\star\frac{\ka}{\Delta_{+}}\pd^{j}\psi^{T}+\frac{1}{4}\psi^{T}\gamma^{jT}\gamma^{iT}\right)
\end{align}
To compute the charge we plug this variation into the first term of \eqref{eq:noether-k3-2} and integrate  over space, which gives us the following expression for the rotation charge:
\begin{align}
\mathcal{M}^{k}_\kappa=&\;-\int d^{3}x\left(\delta_{M_k}\psi^{T}\star\Pi^{0}\right)=-\int d^{3}x\,\epsilon^{ijk}\left[x^{i}\star\frac{\ka}{\Delta_{+}}\pd^{j}\psi^{T}\star\Pi^{0}+\frac{1}{4}\psi^{T}\gamma^{jT}\gamma^{iT}\star\Pi^{0}\right]\nonumber\\
=&\;-i\int d^{3}x\frac{d^{3}p}{\sqrt{2\op}}\frac{d^{3}q}{\sqrt{2\oq}q_{4}/\ka}\epsilon^{ijk}x^{i}\star\frac{\ka}{\Delta_{+}}\pd^{j}\left[u_{s}^{T}(\pb)a_{\pb}^{s}e^{-ipx}+v_{s}^{T}(-S(\pb))b_{\pb}^{\dagger s}e^{-iS(p)x}\right]\nonumber\\
&\star\left[u_{s'}^{*}(\qb)a_{\qb}^{\dagger s'}e^{-iS(q)x}+v_{s'}^{*}(-S(\qb))b_{\qb}^{s'}e^{-iqx}\right]\nonumber\\
&-\frac{i}{4}\int d^{3}x\frac{d^{3}p}{\sqrt{2\op}}\frac{d^{3}q}{\sqrt{2\oq}q_{4}/\ka}\epsilon^{ijk}\left[u_{s}^{T}(\pb)a_{\pb}^{s}e^{-ipx}+v_{s}^{T}(-S(\pb))b_{\pb}^{\dagger s}e^{-iS(p)x}\right]\gamma^{jT}\gamma^{iT}\nonumber\\
&\star\left[u_{s'}^{*}(\qb)a_{\qb}^{\dagger s'}e^{-iS(q)x}+v_{s'}^{*}(-S(\qb))b_{\qb}^{s'}e^{-iqx}\right]
\end{align}
After some tedious manipulations (for details on the calculation method, see Appendix~\ref{AppB}), we obtain
  \begin{align}
\mathcal{M}^{k}_\kappa=\int\frac{d^{3}p}{p_{4}/\ka}\left[i\epsilon^{ijk}\left(\frac{p_{+}^{3}}{\ka^{3}}p^{i}a_{\pb}^{\dagger s}\pdpd{a_{\pb}^{s}}{p_{j}}+S(p^{i})b_{\pb}^{\dagger s}\pdpd{b_{\pb}^{s}}{S(p_{j})}\right)+\frac{1}{2}\sigma_{ss'}^{k}\left(\frac{p_{+}^{3}}{\ka^{3}}a_{\pb}^{\dagger s}a_{\pb}^{s'}+b_{\pb}^{\dagger s}b_{\pb}^{ s'}\right)\right]\label{eq:rot-charge-k3}
\end{align}

\noindent
where $\sigma_{ss'}^{k}$ denotes the elements of  Pauli  matrix $\sigma^k$. Notice that the rotation charges are time-independent, which proves the rotational invariance of the theory.

\subsubsection{Boost charge}

In the case of boosts, we have the transformation
\begin{align}
\delta_{N_j}\psi^{T}=\left(-x^{0}\star\frac{\ka}{\Delta_{+}}\pd^{j}\psi^{T}+x^{j}\star\frac{\ka}{\Delta_{+}}\pd^{0}\psi^{T}+\frac{1}{2}\psi^{T}\gamma^{jT}\gamma^{0T}\right)
\label{eq:varpsi-n-k}
\end{align}
Plugging into \eqref{eq:noether-k3-2}, we obtain the following expression for the boost charge:
\begin{align}
\mathcal{N}^{j}_\kappa=&\;i\int d^{3}x\frac{d^{3}p}{\sqrt{2\op}}\frac{d^{3}q}{\sqrt{2\oq}q_{4}/\ka}x^{[0}\star\frac{\ka}{\Delta_{+}}\pd^{j]}\left[u_{s}^{T}(\pb)a_{\pb}^{s}e^{-ipx}+v_{s}^{T}(-S(\pb))b_{\pb}^{\dagger s}e^{-iS(p)x}\right]\nonumber\\
&\star\left[u_{s'}^{*}(\qb)a_{\qb}^{\dagger s'}e^{-iS(q)x}+v_{s'}^{*}(-S(\qb))b_{\qb}^{s'}e^{-iqx}\right]\nonumber\\
&-\frac{i}{2}\int d^{3}x\frac{d^{3}p}{\sqrt{2\op}}\frac{d^{3}q}{\sqrt{2\oq}q_{4}/\ka}\left[u_{s}^{T}(\pb)a_{\pb}^{s}e^{-ipx}+v_{s}^{T}(-S(\pb))b_{\pb}^{\dagger s}e^{-iS(p)x}\right]\gamma^{jT}\gamma^{0T}\nonumber\\
&\star\left[u_{s'}^{*}(\qb)a_{\qb}^{\dagger s'}e^{-iS(q)x}+v_{s'}^{*}(-S(\qb))b_{\qb}^{s'}e^{-iqx}\right]
\end{align}

As before we present here only the final result, referring the reader to Appendix \ref{AppB} where the boost calculation is presented in detail.

  \begin{align}
\mathcal{N}^j_\kappa=&\;-\frac{i}{2}\int\frac{d^{3}p}{p_{4}/\ka}\frac{p_{+}^{3}}{\ka^{3}}\,\op\left(a_{\pb}^{\dagger s}\pdpd{a_{\pb}^{s}}{p_{j}}-\pdpd{a_{\pb}^{\dagger s}}{p_{j}}a_{\pb}^{s}+3\frac{p^{j}}{p_{+}\op}a_{\pb}^{\dagger s}a_{\pb}^{s}\right)\nonumber\\
&+\frac{i}{2}\int\frac{d^{3}p}{p_{4}/\ka}\osp\left(b_{\pb}^{\dagger s}\pdpd{b_{\pb}^{s}}{S(p_{j})}-\pdpd{b_{\pb}^{\dagger s}}{S(p_{j})}b_{\pb}^{s}+3\frac{p_{+}}{\ka^{2}}\frac{S(p^{j})}{\osp}b_{\pb}^{\dagger s}b_{\pb}^{s}\right)\nonumber\\
&+\frac{1}{2}\int\frac{d^{3}p}{p_{4}/\ka}\,\epsilon^{jkl}p^{k}\sigma^{l}_{ss'}\left(\frac{p_{+}^{3}}{\ka^{3}}\frac{a_{\pb}^{\dagger s}a_{\pb}^{s'}}{\op+m}+\frac{\ka }{p_+}\frac{b_{\pb}^{\dagger s}b_{\pb}^{ s'}}{\osp+m}\right)\label{defboostcharge}
\end{align}

\medskip

The time independence of the boost charge confirms the Lorentz invariance of the theory.
Note that the additional terms proportional to $p^j$ are purely imaginary. They appear as a result of integration by parts, and are an artifact of the chosen normalization of creation and annihilation operators, which can in principle be removed through a different choice of normalization. As such, they are of no significance for the theory.

This completes our calculation of the Noether charges of the $\kappa$-Dirac theory associated with $\kappa$-deformed Poincar\'e symmetries.


\subsection{Symplectic form and charge algebra}
\label{sec:symplecticc form}
In order to obtain the symplectic form, we only need to take the second variation of the time component of the presymplectic current, which we have already calculated:
\begin{align}
\Omega=&\;-\int_{\Sigma_t}d^3x\delta\left(\delta \psi^T\star\Pi^0 \right)=-i\int_{\Sigma_{t}}d^{3}x\left(\delta\psi^{T}\wedge\frac{i\pd_{4}+\ka}{\ka}\star\delta\psi^{*}\right)\nonumber\\
=&\;-i\int_{\Sigma_{t}}\frac{d^{3}p}{p_{4}/\ka}\left[\frac{p_{+}^{3}}{\ka^{3}}\delta a_{\pb}^{s}\wedge\delta a_{\qb}^{\dagger s'}+\delta b_{\pb}^{\dagger s}\wedge\delta b_{\qb}^{s'}\right]
\end{align}
so we have the following Poisson brackets:
\begin{align}
\left\{a_\pb^s,a_\qb^{\dagger s'}\right\}=-i\frac{\ka^3}{p_+^3}\frac{p_4}{\ka}\delta^3(\pb-\qb)\delta^{ss'}\qquad\left\{b_\pb^s,b_\qb^{\dagger s'}\right\}=-i\frac{p_4}{\ka}\delta^3(\pb-\qb)\delta^{ss'}
\label{Poisson brackets}
\end{align}

The brackets have the same expression as the ones that were calculated in our previous work~\cite{CPTscalar}. This is not surprsing, as the action reproduces the same structure in terms of the ordering of the noncommutative fields.

Using the brackets~(\ref{Poisson brackets}), and the expressions (\ref{momdef}), (\ref{eq:rot-charge-k3}), and (\ref{defboostcharge}), one can compute the Poisson brackets between the Poincar\'e charges.
Performing the calculation, it can be shown\footnote{The calculation is very similar to the undeformed case, (see~\cite{CPTscalar} for further details).} that the charges close exactly the standard Poincar\'e algebra, as expected in the classical basis of $\kappa$-Poincar\'e.
This confirms the consistency of the framework also in the case of the fermionic field action.

\section{Discrete symmetries}\label{sec:cpt-both}
In the following Section, we consider the discrete symmetries of the action~\eqref{defLDirac}, as well as its conserved charges obtained in Section~\ref{sec:deformed spin theory}. Given the ambiguity of spacetime inversions in noncommutative spacetime (see for example~\cite{Arzano:2016egk},~\cite{Arzano:2020jro},~\cite{CPTscalar} for three inequivalent formulations), we will examine two prescriptions. In Subsection~\ref{sec:cpt}, we consider the ``minimally deformed'' prescription, in which all spacetime definitions of discrete symmetry transformations are identical to their undeformed counterparts. This approach results in $\mathcal P$- and $\mathcal T$-invariance of the action, but $\mathcal{CPT}$ symmetry is broken due to the inherent violation of $\mathcal C$-symmetry in the deformed theory.  In Subsection~\ref{sec:cpt2}, we utilize the deformed time reversal transformation introduced in~\cite{CPTscalar}, which restores $\mathcal{CPT}$-invariance, but at the cost of $\mathcal T$-invariance.
Since, as we will see, discrete transformations can take us between different Lagrangians and actions, from now on we will explicitly denote $\mathcal S_3^\ka=\int d^4x\;\mathcal L_3^\ka$ rather than simply $\mathcal S^\ka$.
\subsection{Discrete symmetries I}\label{sec:cpt}
In this Subsection we examine the discrete symmetry transformations of the action $\mathcal S_3^\ka$ that result from directly applying the standard spacetime definitions of $\mathcal C$, $\mathcal P$ and $\mathcal T$ to the deformed theory. This conceptually straightforward approach retains $\mathcal P$- and $\mathcal T$-symmetry, but the asymmetry between particle and antiparticle momentum spaces results in $\mathcal C$, and by extension $\mathcal {CPT}$-violation.
\subsubsection{Charge conjugation}
We define the action of charge conjugation on the deformed Dirac field as 
\begin{equation}
\mathcal{C}^\dagger\psi(t,\xb)\mathcal{C}=\eta_C C\bar\psi^T(t,\xb)\qquad \mathcal{C}^\dagger\bar\psi(t,\xb)\mathcal{C}=\eta_C^*\psi^T(t,\xb)C^\dagger\label{eq:c-def}
\end{equation}
where $\eta_C$ is an overall phase we will subsequently ignore and $C$ is a fixed matrix, which we take to be the standard choice $C=i\gamma^2\gamma^0$. The above definition should be understood off-shell, since, as we will now show, the action $\mathcal S_3^\ka$ is not invariant under this transformation and applying it changes the field equations, which we will now demonstrate. Acting with~\eqref{eq:c-def} on the action ~\eqref{defLDirac} we obtain
\begin{align}
\mathcal{C}^{-1}\mathcal S^\ka_3 \mathcal C&=\mathcal C^{-1}\int d^4x\;\psi^{T}\left(i\gamma^{\mu T}\overleftarrow{\pd_{\mu}}-m\right)\star\bar{\psi}^{T}\mathcal C =\int d^4x\bar\psi C^{T}(i\gamma^{\mu T}\overleftarrow{\partial_\mu}-m)\star C^*\psi \nonumber \\
&=\int\;d^4x\;\bar\psi(-i\gamma^\mu\overleftarrow{\partial_\mu}-m)\star\psi\label{eq:s3c}
\end{align}
This result is not equivalent to $\mathcal S_3^\ka$, since the ordering in the action is meaningful. Up to integration by parts,~\eqref{eq:s3c} is actually equivalent to $\mathcal S^\ka_2=\int d^4x\;\mathcal L_2^\ka$ (cf.~\eqref{k-lagrangians}). Through analysis completely analogous to~\eqref{eq:vars-k3}, one can show that for $\mathcal S^\ka_2$ we have the equation of motion
\begin{equation}
\left(-i{S(\slashed{\partial})}-m\right)\psi=0
\end{equation}
which leads to a different mode expansion for $\psi$ (which we will here denote $\psi_{\mathcal C}$ to stress the distinction) compared to~\eqref{eq:modexp-psi-k3}:
\begin{align}
\psi_{\mathcal C}(x)
&=\int\frac{d^{3}p}{\sqrt{2\op}p_{4}/\ka}\left[u_s(-S(\pb )) a^s_\pb e^{-ipx}+v_s(\pb )b^{\dagger s}_{\pb}e^{-iS(p)x}\right]\label{eq:modexp-psi-k2}
\end{align}
Taking this change of mode expansion into account, we derive the action of $\mathcal C$ on field modes:
\begin{align}
\mathcal C^{-1}\psi\mathcal C=C\bar\psi^T_{\mathcal C}&=\int\frac{d^{3}p}{\sqrt{2\op}p_{4}/\ka}\left[C\bar u^T_s(-S(\pb )) a^{\dagger s}_\pb e^{-iS(p)x}+C\bar v^T_s(\pb )b^{ s}_{\pb}e^{-ipx}\right]\nonumber\\
&=\int\frac{d^{3}p}{\sqrt{2\op}p_{4}/\ka}\left[v_s(-S(\pb )) a^{\dagger s}_\pb e^{-iS(p)x}+u_s(\pb )b^{ s}_{\pb}e^{-ipx}\right],
\end{align}
thus
\begin{equation}
\mathcal C^{-1}b_\pb^{\dagger s}\mathcal C=a^{\dagger s}_\pb,\qquad\mathcal C^{-1}a_\pb^s\mathcal C=b_\pb^s.
\end{equation}

Comparing to~\eqref{eq:modexp-psi-k3} we find that despite the broken symmetry, the action of $\mathcal C$ on field modes turns out to also be standard (one can easily show the analogous property for $\bar\psi$).

It is worth mentioning that Noether analysis can be carried out for $\mathcal L^\ka_2$ using the exact same methods as for $\mathcal L^\ka_3$, and the resulting conserved charges are indeed the same as~\eqref{momdef},~\eqref{eq:rot-charge-k3} and~\eqref{defboostcharge}, but with particle and antiparticle modes interchanged. This shows that the $\mathcal C$-relation between $\mathcal L^\ka_3$ and $\mathcal L^\ka_2$ is consistent not only at the level of the action, but also its conserved charges.  Moreover, through analogous manipulations one can show also that $\mathcal C^{-1}\mathcal S_2^\ka\mathcal C\simeq\mathcal S^\ka_3$, where $\simeq$ denotes equivalence up to surface terms.
\subsubsection{Parity}
Following the same approach as for $\mathcal C$, we define the action of $\mathcal P$ in the standard way
\begin{equation}
    \mathcal{P}^\dagger\psi(t,\xb)\mathcal{P}=\eta_P P\psi(t,-\xb)\qquad \mathcal{P}^\dagger\bar\psi(t,\xb)\mathcal{P}=\eta_P^*\bar\psi(t,-\xb)P^\dagger
\end{equation}
where $P=\gamma^0$. We find that the mechanism of $\mathcal P$-symmetry is not affected by the deformation, and the action is easily shown to be invariant:
\begin{align}
    \mathcal{P}^{-1}\mathcal{S}^\ka_3\mathcal P&=\int dtd^{3}x\psi^T(t,-\xb)P^{\dagger}(i\gamma^{\mu T}\overleftarrow{\pd_{\mu}}-m)\star P\bar\psi^T(t,-\xb)\nonumber\\
&=\int dtd^{3}x\psi^T(t,-\xb)(i\gamma^{0}\pd_{0}-i\gamma^{j}\pd_{j}-m)\star\bar\psi^T(t,-\xb)\nonumber\\
&\xrightarrow[\xb'=-\xb]{}\int dtd^{3}x'\psi^T(t,\xb')(i\gamma^{0}\pd_{0}+i\gamma^{j}\pd'_{j}-m)\star\bar\psi^T(t,\xb')=\mathcal{S}^\ka_3\label{eq:s3-p}
\end{align}
It follows that the action on field modes is also undeformed, i.e. $\mathcal P^{-1}a_\pb^s\mathcal P=a^s_{-\pb}$, and so all Poincar\'e charges behave in the standard way under parity. By the same mechanism as~\eqref{eq:s3-p}, one can also show that similarly $\mathcal P^{-1}\mathcal S_2^\ka \mathcal P=\mathcal S_2^\ka$.
\subsubsection{Time reversal - undeformed}
Lastly, we consider the standard time reversal transformation, which we define to act on the field as
\begin{equation}\label{t-undef}
\mathcal{T}^\dagger\psi(t,\xb)\mathcal{T}=\eta_T T\psi(-t,\xb)\qquad \mathcal{T}^\dagger\bar\psi(t,\xb)\mathcal{T}=\eta_T^*\bar\psi(-t,\xb)T^\dagger,
\end{equation}
\begin{equation}
\mathcal{T}^\dagger i\mathcal{T}=-i
\end{equation}
where we take $T=-\gamma^1\gamma^3$. Much like in the case of parity, the symmetries of the action and charges are not affected by the deformation under this prescription, and the action on field modes is standard, i.e. $\mathcal T^{-1}a^s_\pb \mathcal T=(-i\sigma^2)_{ss'}a^{s'}_{-\pb}$ and similarly for $b_\pb^s$, $a^{\dagger s}_\pb$ and $b^{\dagger s}_\pb$. We sketch the proof of invariance of the action:
\begin{align}
\mathcal T^{-1}\mathcal S^\ka_3\mathcal T&=\int dtd^{3}x\psi^T(-t,\xb){T^{T}}(-i\gamma^{\mu \dagger}\overleftarrow{\partial_{\mu}}-m)\star T\bar\psi^T(-t,\xb)\nonumber\\
&=\int dtd^{3}x\psi^T(-t,\xb)(-i\gamma^{0T}\overleftarrow{\pd_{0}}+i\gamma^{jT}\overleftarrow{\partial_{j}}-m)\star\bar\psi^T(-t,\xb)\nonumber\\
&\xrightarrow[t'=-t]{}\int dt'd^{3}x\psi^T(t',\xb)(i\gamma^{0T}\overleftarrow{\pd'_{0}}+i\gamma^{jT}\overleftarrow{\partial_{j}}-m)\star\bar\psi^T(t',\xb)=\mathcal{S}^\ka_3\label{eq:s3-t}
\end{align}
exactly as in the undeformed theory. Since the momentum space mechanism of action is identical to the undeformed case, the standard sign flips of Poincar\'e charges under $\mathcal T$ follow (see Table~\ref{tab:algebra-undeformed}). The proof that $\mathcal T^{-1}\mathcal S_2^\ka\mathcal T=S_2^\ka$ follows the same essential steps as~\eqref{eq:s3-t}.
\subsubsection{$\mathcal{CPT}$}
Using the results~\eqref{eq:s3c},~\eqref{eq:s3-p} and~\eqref{eq:s3-t}, we find that the combined $\mathcal{CPT}$ transformation acts on $\mathcal S_3^\ka$ as
\begin{align}
\mathcal S_3^\ka\xrightarrow[]{\mathcal C+\text{i.b.p.}}\mathcal S_2^\ka\xrightarrow[]{\mathcal P}\mathcal S_2^\ka\xrightarrow[]{\mathcal T}\mathcal S_2^\ka
\end{align}
where i.b.p. stands for integration by parts assuming suitable boundary conditions, and similarly 
\begin{align}
\mathcal S_2^\ka\xrightarrow[]{\mathcal C+\text{i.b.p.}}\mathcal S_3^\ka\xrightarrow[]{\mathcal P}\mathcal S_3^\ka\xrightarrow[]{\mathcal T}\mathcal S_3^\ka
\end{align}
$\mathcal{CPT}$ symmetry is thus broken for the deformed theory and $\mathcal{CPT}$ becomes a nontrivial involutive transformation that relates $\mathcal S_2^\ka$ and $\mathcal S_3^\ka$.\\
At the level of Noether charges, this transformation effectively corresponds to the swapping of particle and antiparticle momentum spaces (we refer to~\cite{CPTscalar} for a thorough discussion) and is closely related to the antipode operation of the $\ka$-Poincar\'e Hopf algebra. As an example, the spatial translation charge transforms as
\begin{align}
(\mathcal{CPT})^{-1}\int\frac{d^{3}p}{p_{4}/\ka}\left[\frac{p_{+}^{3}}{\ka^{3}}p_{\mu}a_{\pb}^{\dagger s}a_{\pb}^{s}-S(p_{\mu})b_{\pb}^{\dagger s}b_{\pb}^{s}\right]\mathcal{CPT}=\int\frac{d^{3}p}{p_{4}/\ka}\left[-S(p_{\mu})a_{\pb}^{\dagger s}a_{\pb}^{s}+\frac{p_{+}^{3}}{\ka^{3}}p_{\mu}b_{\pb}^{\dagger s}b_{\pb}^{s}\right]
\end{align}
Algebraically, this can be described as an extension of the standard table of sign flips under discrete transformations of Noether charges (Table~\ref{tab:algebra-undeformed}) by a second involution $\mathbb S$ corresponding to the momentum space swap (Table~\ref{tab:algebra-deformed-1}). 
\begin{table}[H]
\centering
\begin{minipage}[t]{0.47\textwidth}
\centering
\begin{tabular}{c | c | c | c | c}
\hline
Charge & $\mathcal C$ & $\mathcal P$ & $\mathcal T$ & $\mathcal{CPT}$\\
\hline
$P_0$ & $+$ & $+$ & $+$ & $+$\\
$P_j$ & $+$ & $-$ & $-$ & $+$\\
$M_j$ & $+$ & $+$ & $-$ & $-$\\
$N_j$ & $+$ & $-$ & $+$ & $-$\\
\hline
\end{tabular}
\caption{Undeformed extended charge algebra}
\label{tab:algebra-undeformed}
\end{minipage}\hfill
\begin{minipage}[t]{0.47\textwidth}
\centering
\begin{tabular}{c | c | c | c | c}
\hline
Charge & $\mathcal C$ & $\mathcal P$ & $\mathcal T$ & $\mathcal{CPT}$\\
\hline
$P_0^\ka$ & $+\mathbb S$ & $+$ & $+$ & $+\mathbb S$\\
$P_j^\ka$ & $+\mathbb S$ & $-$ & $-$ & $+\mathbb S$\\
$M_j^\ka$ & $+\mathbb S$ & $+$ & $-$ & $-\mathbb S$\\
$N_j^\ka$ & $+\mathbb S$ & $-$ & $+$ & $-\mathbb S$\\
\hline
\end{tabular}
\caption{Deformed extended charge algebra}
\label{tab:algebra-deformed-1}
\end{minipage}

\end{table}

\subsection{Deformed discrete symmetries II}\label{sec:cpt2}
We now consider the deformed time reversal transformation introduced in~\cite{CPTscalar}, where the deformation of $\mathcal T$ was motivated by inspecting the structure of the curved $\kappa$-momentum space. It differs from~\eqref{t-undef} in that it also applies a combination of complex conjugation and deformed conjugation to the field, which we can denote $\ddagger$\footnote{This operation was previously denoted $\dagger*$, but since the $\dagger$ and $\ast$ operations are ubiquitous in spinor manipulations, we introduce the new symbol to avoid possible confusion.}:
\begin{equation}
    \mathcal{T}_\ka^{-1}\psi(t,\xb)\mathcal{T}_\ka=\eta_T T\psi^\ddagger(-t,\xb)\qquad \mathcal{T}_\ka^{-1}\bar\psi(t,\xb)\mathcal{T}_\ka=\eta_T^*\bar\psi^\ddagger(-t,\xb)T^\dagger,
\end{equation}The action of $\ddagger$ is defined to be
\begin{align}
\psi(x)&=\int\frac{d^{4}p}{p_{4}/\ka}\tilde{\psi}(p)e^{-ipx}\mapsto\int\frac{d^{4}p}{p_{4}/\ka}\tilde{\psi}(p)e^{iS(p)x}\equiv\psi^\ddagger(x)
\end{align}
Note that ${}^\ddagger\xrightarrow[\ka\rightarrow\infty]{}\operatorname{id}$, thus $\mathcal T_\ka$ becomes the standard $\mathcal T$ transformation in the commutative limit. We now examine how the action $\mathcal S^\ka_3$ transforms under $\mathcal T_\ka$:
\begin{align}
\mathcal T_\ka^{-1}\mathcal{S}^\ka_3\mathcal T_\ka&=\int dtd^{3}x\;\psi^{T\ddagger}(-t,\xb)T^{T}\left(-i\left(\gamma^{\mu}\right)^{T*}\overleftarrow{\partial_{\mu}}-m\right)\star T^*\bar{\psi}^{T\ddagger}(-t,\xb)\label{eq:s3-tk-1}
\end{align}
Redefining $t'=-t$ and using the deformed $\delta$ identity $e^{iS(p)x}\star e^{iqx}=\frac{\ka^3}{q_+^3}\frac{q_4}{\ka}\delta^4(S(p)-S(q))$, we obtain
\begin{align}
\mathcal T_\ka^{-1}\mathcal{S}^\ka_3\mathcal T_\ka&=\int dt'd^{3}x\psi^{T\ddagger}(t',\xb)(i\gamma^{0T}\overleftarrow{\pd'_{0}}+i\gamma^{jT}\overleftarrow{\partial_{j}}-m)\star\bar\psi^{T\ddagger}(t',\xb)\nonumber\\
&=\int d^4x\frac{d^{4}p}{p_{4}/\ka}\frac{d^{4}q}{q_{4}/\ka}\tilde{\psi}^{T}(p)e^{iS(p)x}(-\gamma^{\mu T}S(p_{\mu})-m)\star\gamma^{0T}\tilde{\psi}^{\dagger T}(q)e^{iqx}\nonumber\\
&=\int\frac{d^{4}p}{p_{4}/\ka}\tilde{\psi}^{T}(p)(-\gamma^{\mu T}S(p_{\mu})-m)\gamma^{0T}\tilde{\psi}^{\dagger T}(p)
\end{align}
At this point, we use another deformed $\delta$ identity, namely $\frac{q_+^3}{\ka^3}\frac{q_4}{\ka}\delta^4(p-q)=e^{ipx}\star e^{iS(p)x}$, to put the action back in $\star$-product form, but with conjugated plane waves:
\begin{align}
T_\ka^{-1}\mathcal{S}^\ka_3\mathcal T_\ka&=\int\frac{d^{4}p}{p_{4}/\ka}d^{4}q\;\tilde{\psi}^{T}(p)(-\gamma^{\mu T}S(p_{\mu})-m)\gamma^{0T}\tilde{\psi}^{\dagger T}(q)\delta^{4}(p-q)\nonumber\\
&=\int d^4 x \frac{d^{4}p}{p_{4}/\ka}\frac{d^{4}q}{q_{4}/\ka}\frac{\ka^{3}}{q_{+}^{3}}\tilde{\psi}^{T}(p)(-\gamma^{\mu T}S(p_{\mu})-m)\gamma^{0T}\tilde{\psi}^{\dagger T}(q)e^{-ipx}\star e^{-iS(q)x}\nonumber\\
&=\int d^{4}x\frac{\ka^{3}}{\Delta_{+}^{3}}\psi^{T}(-i\gamma^{\mu T}S(\overleftarrow{\pd_{\mu}})-m)\star\bar{\psi}^{T}
\end{align}
Here we recognize the twisted cyclicity formula $\int d^4 x f\star g=\int d^4 x\left(\frac{\ka^3 }{\Delta _+^3} g\right)\star f$, which finally gives us
\begin{align}
T_\ka^{-1}\mathcal{S}^\ka_3\mathcal T_\ka=\int d^{4}x\;\bar{\psi}\star(-i\gamma^{\mu}S(\pd_{\mu})-m)\psi=\mathcal S^\ka_2\label{tk-action}
\end{align}
As we can see, much like $\mathcal C$, $\mathcal T_\ka$  transforms $\mathcal S_3^\ka$ into $\mathcal S_2^\ka$ rather than leave it invariant, and so it is not a symmetry of the deformed action. A proof completely analogous to~\eqref{eq:s3-tk-1}-~\eqref{tk-action} can be constructed to show also that $\mathcal T_\ka^{-1}\mathcal S_2^\ka\mathcal T=\mathcal S_3^\ka$.

For the conserved charges, the action of $\mathcal T_\ka$ is effectively the same as that of $\mathcal C\mathcal T$ (see Table~\ref{tab:algebra-deformed-1}),  although it does not exchange particles and antiparticles, but rather transforms their respective momentum spaces into each other through the $\mathbb S$ operation. Although an expression of $\mathcal T_\ka$ in terms of its action on $a_\pb^s$ and $b_\pb^s$ is possible, it is rather unwieldy and unilluminating. The reason for this is that on-shell, the antipode operation on momenta is no longer an involution, which results in a very cumbersome expression for the following, rather simple, idea:
\begin{equation}
\mathcal T_\ka^{-1}\int\frac{d^{3}p}{p_{4}/\ka}\frac{p_{+}^{3}}{\ka^{3}}p_{j}a_{\pb}^{\dagger s}a_{\pb}^{s}\mathcal T_\ka=\int\frac{d^{3}S(p)}{p_{4}/\ka}\frac{p_{+}^{3}}{\ka^{3}}S(p_{j})a_{\pb}^{\dagger s}a_{\pb}^{s}
\end{equation}
\begin{equation}
\mathcal{T}_\ka^{-1} i\mathcal{T}_\ka=-i
\end{equation}
which would be realized in the undeformed theory as
\begin{equation}
\mathcal{T}^{-1}\int d^{3}p\;p_{j}a_{\pb}^{\dagger s}a_{\pb}^{s}\mathcal{T}=\int d^{3}p\;p_{j}a_{-\pb}^{\dagger s}a_{-\pb}^{s}\xrightarrow[\pb'=-\pb]{}-\int d^{3}p'p'_{j}a_{\pb'}^{\dagger s}a_{\pb'}^{s}
\end{equation}

\subsubsection{$\mathcal{CPT}_\ka$}
Finally, we examine the combination of $\mathcal C$ and $\mathcal P$ with $\mathcal T_\ka$. While both $\mathcal C$ and $\mathcal T_\ka$ symmetries are broken, the order-reversing property of $\mathcal T_\ka$ demonstrated in~\eqref{tk-action} takes the $\mathcal S^\ka_2$ action resulting from $(\mathcal {CP})^{-1}\mathcal S^\ka_3(\mathcal {CP})$ to
\begin{equation}
\mathcal T_\ka^{-1}\int\;d^4x\;\bar\psi(-i\gamma^\mu\overleftarrow{\partial_\mu}-m)\star\psi\mathcal T_\ka=\int d^4x\;\psi^{T}\star\left(-i\gamma^{\mu T}S({\pd_{\mu}})-m\right)\bar{\psi}^{T}
\end{equation}
which is equivalent to $\mathcal S^\ka_3$ up to integration by parts\footnote{Note that  the undeformed Dirac action is also only $\mathcal{CPT}$-invariant up to integration by parts.}.

The action of discrete symmetry transformations on conserved charges using $\mathcal T_\ka$ instead of $\mathcal T$ can be summed up by the following table:
\begin{table}[H]
\centering
\begin{tabular}{c | c | c | c | c}
\hline
Charge & $\mathcal C$ & $\mathcal P$ & $\mathcal T_\ka$ & $\mathcal{CPT}_\ka$\\
\hline
$P_0^\ka$ & $+\mathbb S$ & $+$ & $+\mathbb S$ & $+$\\
$P_j^\ka$ & $+\mathbb S$ & $-$ & $-\mathbb S$ & $+$\\
$M_j^\ka$ & $+\mathbb S$ & $+$ & $-\mathbb S$ & $-$\\
$N_j^\ka$ & $+\mathbb S$ & $-$ & $+\mathbb S$ & $-$\\
\hline
\end{tabular}
\caption{Deformed extended charge algebra with deformed time reversal}
\label{tab:algebra-deformed-2}
\end{table}
The two approaches to discrete symmetries are thus phenomenologically distinct since both predict fundamental $\mathcal C$-breaking, but they differ in whether $\mathcal T$ and $\mathcal {CPT}$ are broken or preserved: the second approach, in which $\mathcal T_\ka$ is deformed, preserves the overall $\mathcal {CPT_\ka}$ symmetry as it was proven in~\cite{CPTscalar} for the scalar field theory.
\section{Conclusions}
\label{sec:conclusions}

In this work we have defined a $\kappa$-Poincar\'e invariant Lagrangian for the free Dirac field and derived its conserved Noether charges associated with spacetime symmetries. The charges satisfy the standard Poincar\'e algebra, as expected in the classical basis of $\kappa$-Poincar\'e. In contrast to the scalar field theory, the first order Dirac operator turns out to have a natural preference for a particular ordering in the action, which sheds some new light on the Hopf-algebra structure of derivatives.

The results of our analysis exhibit the same asymmetric (with respect to particles and antiparticles) momentum structure that was previously (\cite{Arzano:2020jro},~\cite{Bevilacqua:2022fbz},~\cite{CPTscalar}) found for complex scalar fields, specifically, for our choice of Lagrangian they correspond to those for a scalar field described by
\begin{equation}
\mathcal{L}=-\partial_\mu\phi\star S(\pd^\mu)\phi^\dagger-m^2\phi\star\phi^\dagger
\end{equation}
(for details, see \cite{CPTscalar}), and this agreement points towards the self-consistency of the $\kappa$-field theory.

Moreover, the Noether charges we obtained follow the same pattern of $\mathcal C$-symmetry breaking, and the deformed time reversal transformation proposed in~\cite{CPTscalar} can also be applied to Dirac spinors.
In that approach, the overall $\mathcal {CPT}$ is preserved despite $\mathcal{C}$ and $\mathcal{T}$  being broken individually.
This result, here confirmed for the fermion field, is conceptually intriguing in the context of symmetry deformation. Since $\mathcal{CPT}$ symmetry lies at the core of special relativity, its violation would be alarming in a framework in which Planck-scale effects are expected to modify the standard spacetime arena of the theory while preserving its relativistic character, unlike scenarios in which relativistic symmetries are expected to break down at the Planck scale (see \cite{Amelino-Camelia:2008aez,Addazi:2021xuf}).

On the phenomenological side, it would be extremely interesting to look for possible ways of testing the different descriptions of discrete symmetries obtained in our framework, as it was suggested preliminarly in~\cite{CPTscalar}.
The succesful construction of a model for fermions based on the same premises provides a substantial improvement in this direction. 

It is reasonable to expect that these different descriptions lead to markedly different predictions for the outcomes of certain single-particle processes.
A possible way to observe deformation effects for one-particle fermion states is provided by the helicity, defined as the sign of the projection of the spin $\vec s$ onto the direction of the particle momentum $\vec p$,
\begin{equation}
H=\frac{\vec p}{|\vec p|}\cdot\vec{s}.
\label{helicity}
\end{equation}
For a massive particle, the helicity (\ref{helicity}) may change sign under a sufficiently large boost. Intuitively, this is easy to understand, since one can always find a boosted reference frame in which the particle momentum reverses direction with respect to the spin.

For a particle of mass $m$ and energy $E$, the corresponding boost factor\footnote{Notice that in a framework with deformed symmetries, the definition of the boost factor coming from time dilation differs in general with the one defined as the ratio $E/m$, which is the one needed in here. See for instance~\cite{Amelino-Camelia:2025ask}.} is simply $\gamma_0=E/m$.
This effect could in principle be observed for a particle-antiparticle pair produced with nearly equal momenta in the rest frame of a decaying parent particle, such as in the processes $J/\psi\rightarrow e^+e^-$ or $J/\psi\rightarrow\mu^+\mu^-$.
When boosted slightly above $\gamma_0$, the electron (muon) acquires an undeformed energy $E$ and a flipped helicity $-H$, whereas the positron (antimuon) has an energy smaller by $\Delta E=|\vec p|^2/\kappa$ and retains the original helicity $H$.
In this approach, using the particle-antiparticle pair, we can examine a possible deformation of the $\mathcal{CPT}$ transformation since the correct definition of the antiparticle state in the relativistic case requires it to be the $\mathcal{CPT}$ image of that of the particle.
Therefore the scenario \cite{CPTscalar} where the $\mathcal{CPT}$ stays undeformed, and $\mathcal{C}$ and $\mathcal{T}$ are deformed separately and compensating each other, may be insensitive to this kind of test.

From the experimental perspective, however, observing such an effect would require measuring the particle energies with an accuracy of order $\Delta E$, as well as determining the polarizations of both particles.
The required energy resolution for relativistic particles remains orders of magnitude beyond current experimental capabilities.
The polarization of high-energy electrons or muons can nevertheless be measured using two established methods. The first relies on spin asymmetries in elastic scattering off polarized magnetic targets \cite{smc1}, typically achieving an uncertainty at the level of $\sim 2\%$. The second, applicable in the case of muons, makes use of the energy spectrum of electrons from the decay $\mu\rightarrow e\bar\nu_e\nu_{\mu}$ (the Michel spectrum), and also provides a percent-level accuracy \cite{smc2}.

We aim to discuss the phenomenological aspects of the $\kappa$-deformed Dirac field in detail, together with a survey of possible experimental opportunities, in an upcoming publication.

\section*{Acknowledgments}
This work falls within the
scopes of the EU COST Action CA23130 “Bridging high and low energies in search of quantum gravity (BridgeQG)”, and was supported by a STSM Grant from the same COST Action.
For AB, JKG, and GR this work was partially supported by funds provided by the National Science Center, project number 2019/33/B/ST2/00050.

\appendix
\section{Star product on $\kappa$-Minkowski space}\label{AppA}

The $\kappa$-Minkowski spacetime is based on the assumption that coordinates follow the Lie-algebra-type commutation relation $[x^0, \mathbf{x}] = \frac{i}{\kappa} \mathbf{x}$, all other commutators being zero. This algebra is usually called $\mathfrak{an}(3)$ algebra. Several approaches can be used in order to concretely study the consequences of this assumption. The most direct one is to find an adequate representation of the commutation relation, and use it in order to introduce the plane waves (which, due to their nature, are also elements of the group $AN(3)$ obtained from the $\mathfrak{an}(3)$ algebra \cite{Kowalski-Glikman:2002eyl}). The natural group structure will then dictate the properties of momentum space.

A natural representation of the $\mathfrak{an}(3)$ algebra is given by
\begin{equation}\label{x}
	\hat x^0 = -\frac{i}{\kappa} \,\left(\begin{array}{ccc}
		0 & \mathbf{0} & 1 \\
		\mathbf{0}^T & \tilde{\mathbf{0}} & \mathbf{0}^T \\
		1 & \mathbf{0} & 0
	\end{array}\right) \quad
	\hat{x}^i = \frac{i}{\kappa} \,\left(\begin{array}{ccc}
			0 & {(\epsilon^i)\,{}^T} &  0\\
			\epsilon^i & \tilde{\mathbf{0}} & \epsilon^i \\
			0 & -(\epsilon^i)\,{}^T & 0
		\end{array}\right),
\end{equation}
where $(\epsilon^1)^T = (1,0,0)$, $(\epsilon^2)^T = (0,1,0)$, $(\epsilon^3)^T = (0,0,1)$, $\mathbf{0} = (0,0,0)$, and $\tilde{\mathbf{0}}$ is a $3 \times 3$ null matrix. In order to introduce plane waves, because of the non-commutative nature of $x^0$ and $\mathbf{x}$, we use the so called time-to-the-right convention, so that plane waves (and therefore group elements) are defined as
\begin{align}
    \hat{e}_k := e^{i\mathbf{k}\hat{\mathbf{x}}} e^{i k_0 \hat{x}^0}.
\end{align}
Notice that, due to the dimensionful nature of $\hat{x}^0, \hat{\mathbf{x}}$, the quantities $k_0, \mathbf{k}$ have the dimension of momentum, and can therefore be interpreted as coordinates in momentum space. These particular coordinates naturally correspond to translation generators that form what is called the bicrossproduct basis of momentum space~\cite{Majid:1994cy,Agostini:2003vg}. A straightforward calculation allows one to show that
\begin{align}\label{ek}
	\hat{e}_k
	=
	\begin{pmatrix}
		\cosh  \frac{k_0}{\kappa} + \frac{\mathbf{k}^2}{2\kappa^2}e^{\frac{k_0}{\kappa}} & \frac{\mathbf{k}^T}{\kappa} & \sinh \frac{k_0}{\kappa} + \frac{\mathbf{k}^2}{2\kappa^2} e^{\frac{k_0}{\kappa}} \\
		\frac{\mathbf{k}}{\kappa} e^{\frac{k_0}{\kappa}} & \bm{1} & \frac{\mathbf{k}}{\kappa}e^{\frac{k_0}{\kappa}} \\
		\sinh \frac{k_0}{\kappa} - \frac{\mathbf{k}^2}{2\kappa^2}e^{\frac{k_0}{\kappa}} & -\frac{\mathbf{k}^T}{\kappa} & \cosh \frac{k_0}{\kappa} - \frac{\mathbf{k}^2}{2\kappa^2} e^{\frac{k_0}{\kappa}}
	\end{pmatrix}.
\end{align}
\begin{align}\label{groupclassical}
	\hat{e}_{P(k)}
	=
	\frac{1}{\kappa}
	\begin{pmatrix}
		\tilde{P}_4 & \kappa\mathbf{P}/P_+ & P_0 \\
		\mathbf{P} & \kappa \times 1_{3\times 3}&  \mathbf{P} \\
		\tilde{P}_0 & -\kappa\mathbf{P}/P_+ & P_4
	\end{pmatrix}
\end{align}
using the definitions
\begin{eqnarray}\label{classicalbasis2}
	{P_0}(k_0, \mathbf{k}) &=&\kappa  \sinh
	\frac{k_0}{\kappa} + \frac{\mathbf{k}^2}{2\kappa}\,
	e^{  {k_0}/\kappa}, \\
	P_i(k_0, \mathbf{k}) &=&   k_i \, e^{  {k_0}/\kappa},\label{II.1.20}\\
	{P_4}(k_0, \mathbf{k}) &=& \kappa \cosh
	\frac{k_0}{\kappa} - \frac{\mathbf{k}^2}{2\kappa}\, e^{
		{k_0}/\kappa}
\end{eqnarray}
\begin{align}\label{P0eP4tilde}
	\tilde{P}_0
	=
	P_0 - \frac{\mathbf{P}^2}{P_+}
	=
	-S(P_0),
	\qquad
	\tilde{P}_4
	=
	P_4 + \frac{\mathbf{P}^2}{P_+},
    \qquad
    P_+ = P_0 + P_4
\end{align}
These coordinates $P_0, \mathbf{P}, P_4$ satisfy the relations
\begin{align}\label{classicalconstraints}
	-P_0^2 + \mathbf{P}^2 + P_4^2 = \kappa^2
	\qquad
	P_+>0
	\qquad
	P_4>0,
\end{align}
and can be identified with "embedding" coordinates on the (de Sitter) momentum-space hyperboloid.
Furthermore, using the group property, we can define the sum of two different momenta, and the inverse of a momentum by using the following definitions
\begin{align}
    \hat{e}_{P(k)}\hat{e}_{Q(l)} = \hat{e}_{P(k)\oplus Q(l)},
    \qquad
    \hat{e}^{-1}_{P} = \hat{e}_{S(P)}
\end{align}
obtaining
\begin{align}\label{antipode0}
	S(P_0) = -P_0 + \frac{\mathbf{P}^2}{P_0+P_4} = \frac{\kappa^2}{P_0+P_4}-P_4\,,
\end{align}
\begin{align}\label{antipodei4}
	S(\mathbf{P}) =-\frac{\kappa \mathbf{P} }{P_0+P_4}\,,\quad S(P_4) = P_4.
\end{align}
\begin{align}
	(P\oplus Q)_0 &= \frac1\kappa\, P_0(Q_0+Q_4) + \frac{\mathbf{P}\mathbf{Q}}{P_0+P_4} +\frac{\kappa}{P_0+P_4}\, Q_0\label{defsum0-nostar}\\
	(P\oplus Q)_i &=\frac1\kappa\, P_i(Q_0+Q_4) + Q_i\label{defsumi-nostar}\\
	(P\oplus Q)_4 &= \frac1\kappa\, P_4(Q_0+Q_4) - \frac{\mathbf{P}\mathbf{Q}}{P_0+P_4} -\frac{\kappa}{P_0+P_4}\, Q_0\label{defsum4-nostar}\\
    (P\oplus Q)_+&=\frac{1}{\kappa}P_+Q_+\label{defsumplus-nostar}
\end{align}

The classical basis of $\kappa$-Poincar\'e algebra, introduced first in \cite{LukKosMasClassicalBasis}, consists of a redefinition of the translation generators corresponding to the embedding momentum space coordinates.
The peculiarity of the classical basis is that the algebra obeys the standard (undeformed) Poincar\'e commutators (while the co-algebraic sector becomes highly non-trivial, see below).
Moreover, the derivatives corresponding to translation generators in the classical basis, form the differentials of the 5D differential calculus defined in~\cite{Sitarz:1994rh}.

A field theory in $\kappa$-Minkowski can be described in terms of commutative spacetime coordinates through the use of a so-called Weyl map \cite{AgoLizziWeyl2002,Agostini:2003vg}.
Since in this work we define our translation generators to be the ones that comply with the 5D calculus\footnote{An alternative choice that has been pursued in other works (e.g.~\cite{Agostini:2006nc}) is to adopt a 4D calculus~\cite{OecklDiffCalc,Rosati:2021sdf}, so that translation generators are the ones of the bicrossproduct basis \cite{Majid:1994cy}. In that case a natural choice of Weyl map is the one that maps time-ordered plane waves into standard commutative plane waves \cite{Agostini:2003vg}.}, we adopt a particular Weyl map, introduced in \cite{Freidel:2007hk}, with which one can switch from a noncommutative spacetime with coordinates described by the matrices in eq. \eqref{x} and momentum space coordinates $k_\mu$, to a spacetime with commuting coordinates and momentum space coordinates described by the embedding momenta (\ref{classicalbasis2}).
The group structure in this new context manifests itself in the fact that the momenta do not satisfy the canonical addition rules, but the deformed rules in eq. \eqref{defsum0-nostar}, \eqref{defsumi-nostar}, \eqref{defsum4-nostar}. More precisely, the map $\mathcal{W}$ has the definition
\begin{align}
    \mathcal{W}(\hat{e}_k) = e_{P(k)},
    \qquad
    e_P = e^{-iP_0t - i \mathbf{P} \cdot\mathbf{x}}.
\end{align}
\begin{align}
	\mathcal{W}(({\hat{e}  }_k)^{-1}) = e_{S(p  (k))},
	\qquad
	\mathcal{W}(\hat{e}  _k\hat{e}  _l)
	=
	e_{p  (k)\oplus q  (l)} =: e_{p  } \star e_{q  }
\end{align}
The $\star$ product in this context keeps track of the fact that the deformed sum of momenta is non-commutative, so that $f\star g \neq g \star f$. Weyl maps are non-trivial object, with many interesting properties which we will not cover here, more details can be found for example in \cite{AgoLizziWeyl2002,Arzano:2020jro}.

The quantities we have defined until now pertain to the behaviour of a single particle, but the group structure of the $AN(3)$ group is also fundamental in obtaining the superseding Hopf structure, which allows us to deal with multi-particle states (more details can be found in \cite{Majid:1994cy}, \cite{Arzano:2022ewc}, \cite{Kowalski-Glikman:2017ifs}, \cite{Kowalski-Glikman:2002eyl} and references therein). The part of the Hopf algebra dealing with the action on multi-particle states is usually called the co-algebra sector. The determination of the momentum of multi-particle states (or, to use a more concrete notation, the action of momentum operators on the $\star$ product of different functions) is determined by the co-product rules, which in our case turn out to be
\begin{align}\label{DeltaPi}
	\Delta P_i
	=
	\frac{1}{\kappa} P_i \otimes P_+
	+
	\mathbbm{1} \otimes P_i,
\end{align}
\begin{align}\label{DeltaP0}
	\Delta P_0
	=
	\frac{1}{\kappa} P_0 \otimes P_+
	+
	\sum_k
	\frac{P_k}{P_+}
	\otimes
	P_k
	+
	\frac{\kappa}{P_+} \otimes P_0,
\end{align}
\begin{align}\label{DeltaP4}
	\Delta P_4
	=
	\frac{1}{\kappa} P_4 \otimes P_+
	-
	\sum_k
	\frac{P_k}{P_+}
	\otimes
	P_k
	-
	\frac{\kappa}{P_+} \otimes P_0.
\end{align}
The Hopf algebra structure of momentum operators is inherited by partial derivatives. In addition to
\begin{equation}
    \partial_\mu:e^{-ipx}\mapsto -ip_\mu e^{-ipx},
\end{equation}
we also introduce
\begin{align}
\partial_4:e^{-ipx}&\mapsto -i(p_4-\ka)e^{-ipx},\\
S(\partial_\mu):e^{-ipx}&\mapsto -iS(p_\mu)e^{-ipx},\\
\Delta_+:e^{-ipx}&\mapsto p_+e^{-ipx}.
\end{align}

\section{Derivation of deformed Noether charges}\label{AppB}
In this appendix we present the method for computing the conserved Noether charges. Since all of the calculations follow a similar pattern, we will only demonstrate the technique on the most involved example: boosts. To distinguish the nontrivial aspects of this derivation from textbook manipulations, we first review the essential steps of the calculation in the undeformed case, as its results will significantly facilitate the deformed derivation.
The overall formula for the boost charge is, from~\eqref{eq:vars-noether} and~\eqref{eq:varpsi-n},
\begin{align}
\mathcal{N}^j=&\,\,\int d^{3}x\;\psi^{\dagger}\left(-ix^{[0}\pd^{j]}+\frac{i}{2}\gamma^0\gamma^j\right)\psi=\nonumber\\
=&\,\,-\int d^{3}x\frac{d^{3}p}{\sqrt{2\op}}\frac{d^{3}q}{\sqrt{2\oq}}\left(u_{s}^{\dagger}(\pb)a_{\pb}^{\dagger s}e^{ipx}+v_{s}^{\dagger}(\pb)b_{\pb}^{s}e^{-ipx}\right)
(ix^{[0}\pd^{j]})\nonumber\\
&\times\left(u_{s'}(\qb)a_{\qb}^{s'}e^{-iqx}+v_{s'}^{\dagger}(\qb)b_{\qb}^{\dagger s'}e^{iqx}\right)\nonumber\\
&+\frac i 2\int d^{3}x\frac{d^{3}p}{\sqrt{2\op}}\frac{d^{3}q}{\sqrt{2\oq}}\left(u_{s}^{\dagger}(\pb)a_{\pb}^{\dagger s}e^{ipx}+v_{s}^{\dagger}(\pb)b_{\pb}^{s}e^{-ipx}\right)
\gamma^0\gamma^j\nonumber\\
&\times\left(u_{s'}(\qb)a_{\qb}^{s'}e^{-iqx}+v_{s'}^{\dagger}(\qb)b_{\qb}^{\dagger s'}e^{iqx}\right)
\end{align}
We can treat the various combinations of creation and annihilation operators separately, and it suffices to analyze the $a^{\dagger s}_\pb a^{s'}_\qb$ and $a^{\dagger s}_\pb b^{\dagger s'}_\qb$ contributions, which we will label $\mathcal N^j_{aa}$ and $\mathcal N^j_{ab}$ respectively, with the rest following by analogy.
The crucial step in both cases is the Fourier transform of $x^j$ into either $\pdd{p_j}$ or $\pdd{q_j}$, after which we integrate by parts, integrate out the Dirac delta and utilize the $u$ and $v$ spinor properties~\eqref{eq:uv}. For the $\mathcal N^j_{aa}$ term, these steps bring us eventually to
\begin{align}
\mathcal N^j_{aa}=&\,\,i\int d^{3}p\,\left(\frac{1}{2}\frac{p^{j}}{\op}a_{\pb}^{\dagger s}a_{\pb}^{s'}-\op a_{\pb}^{\dagger s}\pdpd{a_{\pb}^{s'}}{p_{j}}\right)-\frac{i}{2}\int d^{3}p\;u_{s}^{\dagger}(\pb)\pdpd{u_{s'}(\pb)}{p_{j}}a_{\pb}^{\dagger s}a_{\pb}^{s'}+\nonumber\\
&+\frac{i}{2}\int\frac{d^{3}p}{2\op}\bar{u}_{s}(\pb)\gamma^{j}u_{s'}(\pb)a_{\pb}^{\dagger s}a_{\pb}^{s'}\label{eq:njaa-1}
\end{align}
Then, using the explicit form~\eqref{eq:uv-def} of $u(\pb)$, through some basic $\gamma$ and $\sigma$ matrix manipulations we find that
\begin{align}
-\frac{i}{2}u_{s}^{\dagger}(\pb)\pdpd{u_{s'}(\pb)}{p_{j}}+\frac{i}{2}\frac{1}{2\op}\bar{u}_{s}(\pb)\gamma^{j}u_{s'}(\pb)=\frac{1}{2}\xi_{s}^{\dagger}\frac{\epsilon^{jkl}p^{k}\sigma^{l}}{\op+m}\xi_{s'}
\end{align}
Integrating the first term in~\eqref{eq:njaa-1} by parts, we obtain
\begin{align}
\mathcal{N}^j_{aa}=-\frac{i}{2}\int d^{3}p\,\op\left(a_{\pb}^{\dagger s}\pdpd{a_{\pb}^{s}}{p_{j}}-\pdpd{a_{\pb}^{\dagger s}}{p_{j}}a_{\pb}^{s}\right)+\frac{1}{2}\int d^{3}p\,\frac{\epsilon^{jkl}p^{k}\sigma^{l}_{ss'}}{\op+m}a_{\pb}^{\dagger s}a_{\pb}^{s'}\label{eq:boost-charge-pt1}
\end{align}
For the $\mathcal{N}^j_{ab}$ contribution, the calculation proceeds similarly, except in this case the Dirac delta $\delta^3(\pb+\qb)$ kills all terms involving the product $u_s^\dagger (\pb )v_{s'}(-\pb )$, leaving us with
\begin{align}
\mathcal{N}^j_{ab}=\frac{i}{2}\int d^{3}p\,\,u_{s}^{\dagger}(\pb)\pdpd{v_{s'}(-\pb)}{p_{j}}e^{2i\op t}a_{\pb}^{\dagger s}b_{-\pb}^{\dagger s'}+\frac{i}{2}\int\frac{d^{3}p}{2\op}\bar{u}_{s}(\pb)\gamma^{j}v_{s'}(-\pb)e^{2i\op t}a_{\pb}^{\dagger s}b_{-\pb}^{\dagger s'}
\label{eq:nj-ab}
\end{align}
which can be shown to vanish using the explicit forms of $u(\pb)$ and $v(\pb)$ and $\gamma$ matrix manipulations. Crucially, the vanishing of $\mathcal N^j_{ab}$ (as well as the other mixed term) ensures the time-independence of the charge.
The remaining terms $\mathcal N^j_{ba}$ and $\mathcal N^j_{bb}$ are evaluated in essentially the same way, resulting in the total charge~\eqref{boost-u}.

In the deformed case, we start from the continuity equation~\eqref{eq:noether-k3-2} and the field variation~\eqref{eq:varpsi-n-k}, which give us the total deformed boost charge
\begin{align}
\mathcal{N}^{j}_\kappa=&\;-\int d^{3}x\left(\delta_{N_j}\psi^{T}\star\Pi^{0}\right)=-\int d^{3}x\left(-x^{[0}\star\frac{\ka}{\Delta_{+}}\pd^{j]}\psi^{T}+\frac{1}{2}\psi^{T}\gamma^{jT}\gamma^{0T}\right)\star\left(i\frac{i\pd_4+\ka}{\ka}\right)\psi^*\nonumber\\
=&\;i\int d^{3}x\frac{d^{3}p}{\sqrt{2\op}p_4/\ka}\frac{d^{3}q}{\sqrt{2\oq}}x^{[0}\star\frac{\ka}{\Delta_{+}}\pd^{j]}\left[u_{s}^{T}(\pb)a_{\pb}^{s}e^{-ipx}+v_{s}^{T}(-S(\pb))b_{\pb}^{\dagger s}e^{-iS(p)x}\right]\nonumber\\
&\star\left[u_{s'}^{*}(\qb)a_{\qb}^{\dagger s'}e^{-iS(q)x}+v_{s'}^{*}(-S(\qb))b_{\qb}^{s'}e^{-iqx}\right]\nonumber\\
&-\frac{i}{2}\int d^{3}x\frac{d^{3}p}{\sqrt{2\op}p_4/\ka}\frac{d^{3}q}{\sqrt{2\oq}}\left[u_{s}^{T}(\pb)a_{\pb}^{s}e^{-ipx}+v_{s}^{T}(-S(\pb))b_{\pb}^{\dagger s}e^{-iS(p)x}\right]\gamma^{jT}\gamma^{0T}\nonumber\\
&\star\left[u_{s'}^{*}(\qb)a_{\qb}^{\dagger s'}e^{-iS(q)x}+v_{s'}^{*}(-S(\qb))b_{\qb}^{s'}e^{-iqx}\right]
\end{align}
Similarly to the undeformed case, we will examine the $\mathcal N^j_{\ka aa}$ and $\mathcal N^j_{\ka ab}$ contributions, starting with $\mathcal N^j_{\ka aa}$:
\begin{align}
\mathcal{N}^{j}_{\kappa aa}=&\;i\int d^{3}x\frac{d^{3}p}{\sqrt{2\op}p_4/\ka}\frac{d^{3}q}{\sqrt{2\oq}}x^{[0}\star\frac{\ka}{\Delta_{+}}\pd^{j]}u_{s}^{T}(\pb)a_{\pb}^{s}e^{-ipx}\star u_{s'}^{*}(\qb)a_{\qb}^{\dagger s'}e^{-iS(q)x}\nonumber\\
&-\frac{i}{2}\int d^{3}x\frac{d^{3}p}{\sqrt{2\op}p_4/\ka}\frac{d^{3}q}{\sqrt{2\oq}}u_{s}^{T}(\pb)a_{\pb}^{s}e^{-ipx}\gamma^{Tj}\gamma^{T0}\star u_{s'}^{*}(\qb)a_{\qb}^{\dagger s'}e^{-iS(q)x}
\end{align}
To get rid of the first $\star$-product, we make use of the following identity
\begin{align}\label{eq:x-star-phi}
x_{\mu}\star\phi(x)=\frac{1}{\ka}\left(x_{\mu}\Delta_{+}-ix_{0}\pd_{\mu}\right)\phi(x)
\end{align}
which was proved in~\cite{CPTscalar}. We also use
\begin{align}
\int d^3x d^3p\;e^{-ipx}\star e^{-iS(q)x}=\int d^3p\;e^{-i(\op\oplus S(\oq))t}\frac{q_+^3}{\ka^3}\delta^3(\pb-\qb)
\end{align}
to integrate out one of the momentum variables in the spin term:
\begin{align}
\mathcal N^j_{\ka aa}=&\;\int d^{3}x\frac{d^{3}p}{\sqrt{2\op}p_4/\ka}\frac{d^{3}q}{\sqrt{2\oq}}\frac{\ka}{p_{+}}\frac{1}{\ka}\left[t\ka\frac{p_{+}}{q_{+}}-t\left(\op\oplus S(\oq)\right)\right]p^{j}u_{s}^{T}(\pb)a_{\pb}^{s}u_{s'}^{*}(\qb)a_{\qb}^{\dagger s'}e^{-i(p\oplus S(q))x}\nonumber\\
&-\int d^{3}x\frac{d^{3}p}{\sqrt{2\op}p_4/\ka}\frac{d^{3}q}{\sqrt{2\oq}}\frac{\ka}{p_{+}}\frac{1}{\ka}\left[x^{j}\ka\frac{p_{+}}{q_{+}}-t\left(p\oplus S(q)^j\right)\right]\op u_{s}^{T}(\pb)a_{\pb}^{s}u_{s'}^{*}(\qb)a_{\qb}^{\dagger s'}e^{-i(p\oplus S(q))x}\nonumber\\
&-\frac{i}{2}\int \frac{d^{3}p}{2\op p_{4}/\ka}\frac{p_{+}^{3}}{\ka^{3}}u_{s}^{T}(\pb)\gamma^{jT}\gamma^{0T}u_{s'}^{*}(\pb)a_{\pb}^{s}a_{\pb}^{\dagger s'}
\end{align}
Since $(p\oplus S(q))^\mu|_{p=q}=0$ by definition, the second terms in brackets can be ignored, as they will eventually be removed by the Dirac delta. The remaining nontrivial step is the Fourier transform of $x^j$. Considering that
\begin{equation}
e^{-i(p\oplus S(q))_jx^j}=e^{-i\frac{\ka}{q_+}(p_j-q_j)x^j}
\end{equation}
is linear in $\pb$ and nonlinear in $\qb$, there is a natural preference to express $x^j$ as
\begin{equation}
x^j e^{-i(p\oplus S(q))x}=e^{-i(\op\oplus S(\oq))t}\left(i\frac{q_+}{\ka}\pdd{p_j}\right)e^{-i\frac{\ka}{q_+}(\pb-\qb)\cdot \xb}
\end{equation}
which brings us to
\begin{align}
\mathcal{N}^j_{\kappa aa}=&\;\int d^{3}x\frac{d^{3}p}{\sqrt{2\op}p_4/\ka}\frac{d^{3}q}{\sqrt{2\oq}}\frac{\ka}{q_{+}}tp^{j}u_{s}^{T}(\pb)a_{\pb}^{s}u_{s'}^{*}(\qb)a_{\qb}^{\dagger s'}e^{-i(p\oplus S(q))x}\nonumber\\
&-\int d^{3}x\frac{d^{3}p}{\sqrt{2\op}p_4/\ka}\frac{d^{3}q}{\sqrt{2\oq}}\frac{\ka}{q_{+}}\op u_{s}^{T}(\pb)a_{\pb}^{s}u_{s'}^{*}(\qb)a_{\qb}^{\dagger s'}e^{-i(\op\oplus S(\oq))t}\left(i\frac{q_{+}}{\ka}\pdd{p_{j}}\right)e^{-i\frac{\ka}{q_{+}}\left(p_{i}-q_{i}\right)x^{i}}\nonumber\\
&-\frac{i}{2}\int \frac{d^{3}p}{2\op p_{4}/\ka}\frac{p_{+}^{3}}{\ka^{3}}u_{s}^{T}(\pb)\gamma^{jT}\gamma^{0T}u_{s'}^{*}(\pb)a_{\pb}^{s}a_{\pb}^{\dagger s'}\nonumber\\
=&\;i\int\frac{d^{3}p}{2\sqrt{\op}p_{4}/\ka}\frac{p_{+}^{3}}{\ka^{3}}\left(\pdd{p_{j}}\sqrt{\op}u_{s}^{T}(\pb)a_{\pb}^{s}-it\pdd{p_{j}}(\op\oplus S(\oq))\bigg|_{\pb=\qb}\right)u_{s'}^{*}(\pb)a_{\pb}^{\dagger s'}\nonumber\\
&+\int\frac{d^{3}p}{p_{4}/\ka}\frac{p_{+}^{3}}{\ka^{3}}\frac{\ka}{p_{+}}tp^{j}a_{\pb}^{s}a_{\pb}^{\dagger s}-\frac{i}{2}\int \frac{d^{3}p}{2\op p_{4}/\ka}\frac{p_{+}^{3}}{\ka^{3}}u_{s}^{T}(\pb)\gamma^{jT}\gamma^{0T}u_{s'}^{*}(\pb)a_{\pb}^{s}a_{\pb}^{\dagger s'}
\end{align}
The time-dependent term evaluates to
\begin{align}
\pdd{p_{j}}(\op\oplus S(\oq))\xrightarrow[]{\qb=\pb}-\frac{\ka}{p_+}\frac{p^j}{\op}
\end{align}
and so it exactly cancels the other time-dependent term, leaving us with
\begin{align}
\mathcal{N}^j_{\kappa aa}=&\;-i\int\frac{d^{3}p}{p_{4}/\ka}\frac{p_{+}^{3}}{\ka^{3}}\left(\frac{1}{2}\frac{p^{j}}{\op}a_{\pb}^{s}-\op\pdpd{a_{\pb}^{s}}{p_{j}}\right)a_{\pb}^{\dagger s}+i\int\frac{d^{3}p}{2\op p_{4}/\ka}\frac{p_{+}^{3}}{\ka^{3}}\pdpd{u_{s}^{T}(\pb)}{p_{j}}u_{s'}^{*}(\pb)a_{\pb}^{s}a_{\pb}^{\dagger s'}\nonumber\\
&-\frac{i}{2}\int \frac{d^{3}p}{2\op p_{4}/\ka}\frac{p_{+}^{3}}{\ka^{3}}u_{s}^{T}(\pb)\gamma^{jT}\gamma^{0T}u_{s'}^{*}(\pb)a_{\pb}^{s}a_{\pb}^{\dagger s'}
\end{align}
It is easy to see that this is exactly the transpose of the analogous expression for the undeformed case~\eqref{eq:njaa-1} multiplied by the factor $\frac{p_+^3}{\ka^3}\frac{\ka}{p_4}$. Because of this additional factor, the first term does not admit the same simplification \eqref{eq:boost-charge-pt1} via integration parts, which results in an extra purely imaginary translation-like term; however, it should be noted that this is of no physical significance, as it is just an artifact of the normalization choice for $a_\pb^s$.
In the end, the $\mathcal{N}^j_{\kappa aa}$ contribution is
\begin{align}
\mathcal{N}^j_{\kappa aa}=-\frac{i}{2}\int\frac{d^{3}p}{p_{4}/\ka}\frac{p_{+}^{3}}{\ka^{3}}\,\op\left(a_{\pb}^{\dagger s}\pdpd{a_{\pb}^{s}}{p_{j}}-\pdpd{a_{\pb}^{\dagger s}}{p_{j}}a_{\pb}^{s}+3\frac{p^{j}}{p_{+}\op}a_{\pb}^{\dagger s}a_{\pb}^{s}\right)+\frac{1}{2}\int\frac{d^{3}p}{p_{4}/\ka}\frac{p_{+}^{3}}{\ka^{3}}\,\frac{\epsilon^{jkl}p^{k}\sigma^{l}_{ss'}}{\op+m}a_{\pb}^{\dagger s}a_{\pb}^{s'}
\end{align}

Due to the transposition of the action, the mixed term analogous to the undeformed $\mathcal N^j_{ab}$ is
\begin{align}
    \mathcal{N}^{j}_{\ka ba}=&\;i\int d^{3}x\frac{d^{3}p}{\sqrt{2\op}}\frac{d^{3}q}{\sqrt{2\oq}q_{4}/\ka}x^{[0}\star\frac{\ka}{\Delta_{+}}\pd^{j]}v_{s}^{T}(-S(\pb))b_{\pb}^{\dagger s}e^{-iS(p)x}\star u_{s'}^{*}(\qb)a_{\qb}^{\dagger s'}e^{-iS(q)x}\nonumber\\
    &-\frac{i}{2}\int d^{3}x\frac{d^{3}p}{\sqrt{2\op}}\frac{d^{3}q}{\sqrt{2\oq}q_{4}/\ka}v_{s}^{T}(-S(\pb))b_{\pb}^{\dagger s}e^{-iS(p)x}\gamma^{jT}\gamma^{0T}\star u_{s'}^{*}(\qb)a_{\qb}^{\dagger s'}e^{-iS(q)x}
\end{align}
Its computation proceeds analogously to $\mathcal N^j_{\ka aa}$. Using
\begin{equation}
\int d^3x\,d^3p\,e^{-iS(p)x}\star e^{-iS(q)x}=\int\,d^3p\,e^{-i(S(\op)\oplus S(\oq))t}\frac{q_+^3}{\ka^3}\delta^3(S(\pb)-\qb)
\end{equation}
we find that all terms involving $v_s^T(-S(\pb))u_{s'}^*(\qb)$ evaluate to \begin{equation}
    \left(u^\dagger_{s'}(\qb)v_{s}(-\qb)\right)^T=0
\end{equation}
The Fourier transform of $x^j$ is in this case done through
\begin{equation}
x^je^{-i(S(p)\oplus S(q))x}=e^{-i(S(\op)t\oplus S(\oq))t}\left(i\frac{q_+}{\ka}\pdd{S(p_j)}\right)e^{-i\frac{\ka}{q_+}(S(\pb)-\qb)\cdot\xb}
\end{equation}
and the only terms surviving $\delta^3(S(\pb)-\qb)$ are
\begin{align}
\mathcal N^j_{\ka ba}=&\;-\frac{i}{2}\int\frac{d^{3}S(p)}{p_{4}/\ka}\sqrt{\frac{\op}{\osp}}\frac{p_{+}^{3}}{\ka^{3}}\frac{p_{S+}^{3}}{\ka^{3}}\pdpd{v_{s}^{T}(-S(\pb))}{S(p_{j})}u_{s'}^{*}(S(\pb))b_{\pb}^{\dagger s}a_{S(\pb)}^{\dagger s'}e^{-i(S(\op)\oplus S(\osp))t}\nonumber\\
&-\frac{i}{2}\int\frac{d^{3}S(p)}{2\osp p_{4}/\ka}\sqrt{\frac{\op}{\osp}}\frac{p_{+}^{3}}{\ka^{3}}\frac{p_{S+}^{3}}{\ka^{3}}v_{s}^{T}(-S(\pb))\gamma^{jT}\gamma^{0T}u_{s'}^{*}(S(\pb))b_{\pb}^{\dagger s}a_{S(\pb)}^{\dagger s'}e^{-i(S(\op)\oplus S(\osp))t}
\label{eq:nj-ba-k}
\end{align}
Although this is less obvious than in the case of $\mathcal N^j_{\ka aa}$,~\eqref{eq:nj-ba-k} is again structurally the same as~\eqref{eq:nj-ab} modulo some $\pb^2$-dependent factors, with the $\pb$ variable replaced by $S(\pb)$. It thus vanishes on the same grounds, and in the end
\begin{equation}
\mathcal{N}^j_{\ka ba}=0
\end{equation}

The remaining terms of the boost charge, as well as all the other conserved charges, can be obtained using the same methods as we delineated above. For details on the relevant formulas, as well as proofs of some of the less obvious properties of the $\star$-product calculus, we refer the reader to the Appendices of~\cite{CPTscalar}.

\end{document}